\documentclass[conference,compsoc]{IEEEtran}

\usepackage{enumerate}
\usepackage{enumitem}
\usepackage{graphicx}
\usepackage{epstopdf}
\usepackage{epsfig}
 \usepackage[hyphens]{url}
\usepackage{etoolbox}
\usepackage{tikz}
\usepackage{amsmath}
\usepackage{stmaryrd}
\usepackage{mathtools}
\usepackage{hhline}
\usepackage{setspace}
\usepackage{booktabs}
\usepackage{multicol}
\usepackage{multirow}
\usepackage{xcolor,colortbl}
\usepackage{array}
\usepackage{xspace}
\usepackage{pifont}
\usepackage{graphicx}
\usepackage{subcaption}
\usepackage[most]{tcolorbox}
\usepackage[htt]{hyphenat}

\usepackage[flushleft]{threeparttable}
\usepackage[multiple]{footmisc}
\usepackage{adjustbox}
\usepackage{diagbox}
\usetikzlibrary{arrows}
\usepackage{wasysym}
\usepackage{longtable}
\usepackage{tabularx}

\usepackage{listings}
\usepackage{color}
\usepackage{enumitem}

\usepackage{fontawesome5}

\usepackage{hyperref}
\hypersetup{
  hidelinks,
	linkcolor = {darkpurple},
  breaklinks = true
}

\usepackage{cleveref}

\definecolor{dkgreen}{rgb}{0,0.6,0}
\definecolor{gray}{rgb}{0.5,0.5,0.5}
\definecolor{mauve}{rgb}{0.58,0,0.82}
\definecolor{Gray}{gray}{0.9}
\definecolor{persiangreen}{rgb}{0.0, 0.65, 0.58}
\definecolor{viridian}{rgb}{0.25, 0.51, 0.43}
\definecolor{ruddy}{rgb}{1.0, 0.0, 0.16}
\definecolor{lilac}{HTML}{AC5FD9}

\newcommand{\xmark}{\color{ruddy}\ding{55}}%
\newcommand{\rrow}{$\rightarrow$}

\newcommand{\nmh}{\texttt{NativeMessaging}\xspace}
\newcommand{\nm}{native messaging\xspace}

\DeclareRobustCommand*\circledchar[2]{
  \tikz[baseline=(c.base)]
    \node[circle, white, font=\sffamily\bfseries, fill={#1},
    draw={#1}, inner sep=0pt, minimum size=1.33em](c){#2};
}

\newcommand*\circleblack[1]{\scalebox{0.8}{\circledchar{black}{#1}}}
\newcommand*\circlepurple[1]{\scalebox{0.8}{\circledchar{lilac}{#1}}}

\definecolor{borderblue}{HTML}{006EAF}
\definecolor{fillblue}{HTML}{1BA1E2}

\usepackage{pgfplots}
\pgfplotsset{compat=1.16}

\definecolor{lightgray}{rgb}{0.95, 0.95, 0.95}
\definecolor{darkgray}{rgb}{0.4, 0.4, 0.4}
\definecolor{purple}{rgb}{0.65, 0.12, 0.82}
\definecolor{editorOcher}{rgb}{1, 0.5, 0} 
\definecolor{editorGreen}{rgb}{0, 0.5, 0} 
\definecolor{orange}{rgb}{1,0.45,0.13}		
\definecolor{olive}{rgb}{0.17,0.59,0.20}
\definecolor{brown}{rgb}{0.69,0.31,0.31}
\definecolor{purple}{rgb}{0.38,0.18,0.81}
\definecolor{lightblue}{rgb}{0.1,0.57,0.7}
\definecolor{lightred}{rgb}{1,0.4,0.5}

\definecolor{darkpurple}{rgb}{0.502, 0.26, 0.747}
\definecolor{background}{HTML}{EEEEEE}
\definecolor{commentgreen}{RGB}{2,112,10}
\colorlet{numb}{magenta!60!black}
\definecolor{eminence}{RGB}{108,48,130}

\usepackage{xurl}
\usepackage{upquote}

\definecolor{eclipseStrings}{RGB}{42,0.0,255}
\definecolor{eclipseKeywords}{RGB}{127,0,85}
\colorlet{numb}{magenta!60!black}
\lstdefinelanguage{json}{
  keywordstyle=\color{blue}\bfseries,
  keywords={permissions, content\_scripts,manifest\_version,background,browser\_action},
  keywordstyle=\color{lightblue},
  identifierstyle=\color{black},
  sensitive=false,
  comment=[l]{//},
  morecomment=[s]{/*}{*/},
  commentstyle=\color{purple}\ttfamily,
  stringstyle=\color{red}\ttfamily,
  frame=single,
    literate=
     *{0}{{{\color{numb}0}}}{1}
      {1}{{{\color{numb}1}}}{1}
      {2}{{{\color{numb}2}}}{1}
      {3}{{{\color{numb}3}}}{1}
      {4}{{{\color{numb}4}}}{1}
      {5}{{{\color{numb}5}}}{1}
      {6}{{{\color{numb}6}}}{1}
      {7}{{{\color{numb}7}}}{1}
      {8}{{{\color{numb}8}}}{1}
      {9}{{{\color{numb}9}}}{1}
      {\{}{{{\color{delim}{\{}}}}{1}
      {\}}{{{\color{delim}{\}}}}}{1}
      {[}{{{\color{delim}{[}}}}{1}
      {]}{{{\color{delim}{]}}}}{1},
}

\definecolor{delim}{RGB}{20,105,176}
\definecolor{numb}{RGB}{106, 109, 32}
\definecolor{string}{rgb}{0.64,0.08,0.08}

\lstdefinelanguage{Swift}{
  keywords={typeof, new, true, false, catch, function, func, async, return, null, catch, switch, var, if, in, while, do, else, case, break, AVCaptureSession, AVCaptureDeviceInput, AVCaptureDevice, video, authorized, notDetermined},
  keywordstyle=\color{blue}\bfseries,
  ndkeywords={class, export, boolean, throw, implements, import, this},
  ndkeywordstyle=\color{darkgray}\bfseries,
  identifierstyle=\color{black},
  sensitive=false,
  comment=[l]{//},
  morecomment=[s]{/*}{*/},
  commentstyle=\color{purple}\ttfamily,
  frame=single,
  stringstyle=\color{red}\ttfamily,
  morestring=[b]',
  morestring=[b]"
}

\lstdefinelanguage{json}{
  keywordstyle=\color{blue}\bfseries,
  keywords={Camera, Data,Type, Purpose, Handling, User, Permission, API, Access, true, false, notdetermined},
  keywordstyle=\color{lightblue},
  identifierstyle=\color{black},
  sensitive=false,
  comment=[l]{//},
  morecomment=[s]{/*}{*/},
  commentstyle=\color{purple}\ttfamily,
  stringstyle=\color{red}\ttfamily,
  frame=single,
    literate=
     *{0}{{{\color{numb}0}}}{1}
      {1}{{{\color{numb}1}}}{1}
      {2}{{{\color{numb}2}}}{1}
      {3}{{{\color{numb}3}}}{1}
      {4}{{{\color{numb}4}}}{1}
      {5}{{{\color{numb}5}}}{1}
      {6}{{{\color{numb}6}}}{1}
      {7}{{{\color{numb}7}}}{1}
      {8}{{{\color{numb}8}}}{1}
      {9}{{{\color{numb}9}}}{1}
      {\{}{{{\color{delim}{\{}}}}{1}
      {\}}{{{\color{delim}{\}}}}}{1}
      {[}{{{\color{delim}{[}}}}{1}
      {]}{{{\color{delim}{]}}}}{1},
}

\usepackage{minted}

\usepackage{listings}
\usepackage{xcolor}

\lstdefinelanguage{json}{
    basicstyle=\ttfamily\small,
    numbers=none,
    breaklines=true,
    showstringspaces=false,
    stringstyle=\color{blue},
    morestring=[b]",
    literate=
     *{:}{{\textcolor{red}{:}}}{1}
      {,}{{\textcolor{red}{,}}}{1}
      {\{}{{\textcolor{red}{\{}}}{1}
      {\}}{{\textcolor{red}{\}}}}{1}
}

\usepackage[T1]{fontenc}
\usepackage{textcomp}

\begin{document}

\title{The Native Menace: Host-to-Browser Cross-Boundary Attacks via Native Messaging}
\author{
\IEEEauthorblockN{Jyotirmay Chauhan\textsuperscript{*}, Rownak Islam\textsuperscript{*}, Jason Polakis}
\IEEEauthorblockA{\textit{University of Illinois Chicago}\\
\{jchauh3, misla84, polakis\}@uic.edu}
}
\maketitle
\begingroup\renewcommand\thefootnote{*}
\footnotetext{These authors contributed equally to this work.}
\endgroup

\begin{abstract}
Modern browsers are complex software platforms
that users can customize through browser extensions to 
further expand their functionality. Extensions derive their
 functionality from privileged browser APIs that expose 
powerful capabilities.
Prior work has extensively
studied how malicious webpages or extensions can target
those capabilities. However, prior research has overlooked the
NativeMessaging API, which allows extensions to create
a communication channel that bridges the gap between web
content rendered in the browser and native apps running on
the user’s machine. This API enables workflows that cross the
browser’s boundary to perform critical operations (e.g., digital
signing) or exchange sensitive data (e.g., credentials). 
In this paper, we conduct the first, to our knowledge, empirical security
analysis of NativeMessaging, and present a novel attack
that allows a non-privileged malicious native app to misuse
browsers’ native messaging capabilities to stealthily intercept
sensitive communications or launch confused deputy attacks
against other native apps. To demonstrate the significant risk
introduced by native messaging, we present a series of concrete
attacks against popular and critical apps (e.g., 1Password, the
European Union’s eID system, Google Drive), whose browser
extensions have more than 532 million cumulative downloads.
Our attacks have severe ramifications, including full password
vault leakage, and arbitrary document signing in governmental
workflows. Fundamentally, these attacks are possible due to
native messaging being built on top of an untrusted communication 
channel without providing a root of trust or enforcing
some form of client attestation. Accordingly, we propose a
straightforward mitigation technique of message encryption
with out-of-band secret sharing, to prevent our attacks.
\end{abstract}

\section{Introduction}
\label{sec:intro}


The Internet has become an integral part of modern everyday life, and web
browsers mediate a broad range of our sensitive online activities, including
inter-personal communications, financial tasks, and business-related
workflows~\cite{ruth2022world}. To support the demands of modern web
apps, browser vendors continuously introduce new APIs that expose
richer functionality to web content~\cite{oh2022deview}, ranging from hardware
access to credential management and local
storage~\cite{westers2024single,snyder2016browser,mobileWebAPIAttacks2019}.  As
a result, contemporary browsers have evolved into highly complex software
platforms~\cite{ryck2014browser} deployed across billions of devices worldwide.
At the same time, browsers support extension ecosystems that allow third-party
developers to further augment browser functionality with customized features,
and tighter integration. 


As such, browser extensions have become deeply embedded in the modern web
ecosystem.  Popular browsers such as Google Chrome and Mozilla
Firefox collectively host hundreds of thousands of publicly available
extensions that serve millions of users~\cite{hsu2024chrome}. These extensions
are routinely granted access to highly sensitive browser
privileges~\cite{bauer2014analyzing,kariryaa2021understanding}, including the
ability to read and modify web page contents, intercept network requests, and
access authentication tokens. While these capabilities enable powerful
functionality, they also significantly expand the browser attack surface by
introducing privileged third-party code into security-sensitive
environments~\cite{carlini2012evaluation,heule2015most}.


Consequently, browser extensions have attracted substantial attention from the
security and privacy research community. Prior work has extensively
investigated malicious extensions~\cite{kapravelos2014hulk,xing2015understanding,pantelaios2020you},
privilege escalation attacks~\cite{kim2023extending}, and extension
fingerprinting~\cite{starov2017xhound,karami2020carnus,agarwal2024peeking,sjosten2021essentialfp,sjosten2019latex}.
Research has also demonstrated how extensions can exfiltrate sensitive browsing
data~\cite{xie2024arcanum}, and bypass isolation
boundaries~\cite{wang2021webenclave}. Other studies have explored the security
implications of extension ecosystems at scale~\cite{190984}, uncovering
widespread overprivilege~\cite{some2019empoweb} and insufficient vetting
practices in extension marketplaces~\cite{moreno2026did}. Despite this extensive body of work,
existing research has focused on threats originating from malicious web content
targeting extensions~\cite{309832} or malicious extensions themselves.


In this paper, we adopt a different threat model: that of malicious native
apps misusing browser extension functionality, which is made possible
by \texttt{NativeMessaging}, an overlooked browser extension API. In more
detail, we conduct the first systematic empirical analysis of the \texttt{NativeMessaging}
API, which bridges the gap between the web content rendered in
browsers and native apps. Specifically, this API allows extensions to
communicate with locally installed native apps through
operating-system-level inter-process communication channels. A representative
use case is that of a native password manager sending the credentials for a
specific domain to its browser extension counterpart, which the extension can
then autofill in the website's login form. This allows the extension to limit
its functionality to interacting with websites, without the need to store the
user's vault locally in any of the browser's storages.  Our in-depth empirical
analysis uncovers a series of inherent flaws in the design and deployment of
\emph{NativeMessaging}, rendering it a powerful and underexplored element of
major browsers' attack surface. 

Guided by our findings, we develop a novel cross-boundary attack wherein a
user-level native app (running without any special privileges or
permissions) can exploit \texttt{NativeMessaging} to intercept
the communication between native apps and
browser extensions or deploy confused deputy
attacks against them~\cite{rajani2016access}. Subsequently, we present a series of use
cases with popular and security-critical apps that concretely
demonstrate the severe impact of this attack vector. Among others, our attacks allow
adversaries to stealthily (i)~impersonate users in government services
(including citizen and judicial portals) across the EU, Latin America, and India,
and sign legally-binding documents, (ii)~extract full password vaults
from password managers, (iii)~access and interfere with video-conferencing sessions,
and (iv)~obtain authentication tokens, and (v)~misuse screen capture and keystroke injection permissions. 
While we focus our analysis on Google Chrome, we have confirmed that Firefox is
also susceptible to attacks.

Overall, our findings expose risks arising from the increasingly tight
integration between browsers and native software ecosystems, with particularly
significant implications on macOS, where extensive app-isolation and data
privacy mechanisms provide strong security boundaries between apps. Our attacks
nevertheless cross these boundaries, demonstrating that Native Messaging can
undermine protections that otherwise isolate apps and their data. This
underscores the need for stronger isolation and authentication mechanisms for
extension-native communication channels. To fully prevent our attacks, we
detail a series of mitigations targeting all entities involved in the Native
Messaging workflow, namely, the OS, browser, and native app.

In summary, our research contributions are:
\begin{itemize}
\item We conduct an in-depth empirical analysis of the
\texttt{NativeMessaging} API, and uncover fundamental design flaws that
expose users to significant risks.

\item We present two variants of our Host-to-Browser Cross-Boundary attack:
\textit{bidirectional interception},
a full man-in-the-middle attack between the extension and native host,
and \textit{unidirectional interception}, a
confused-deputy attack against either the extension \emph{or} the native host.

\item We demonstrate the severe implications of our attack vector in macOS, through
concrete attack instantiations against security-critical and highly-sensitive
extensions used by hundreds of millions of users.

\item We detail the necessary changes required for fortifying native messaging
against cross-boundary attacks. We have also disclosed our findings to the
affected vendors, and have released code demonstrating our attack techniques~\cite{NM-artifacts} 
to facilitate additional research in the space.
\end{itemize}

\section{Background and Threat Model}
\label{sec:background}

Here, we first provide a detailed overview of the
\texttt{NativeMessaging} API and protocol. Subsequently, we present an
empirical analysis of the native messaging workflow, and highlight the flaws
that we have uncovered.


%

\textbf{Browser-to-OS boundary.} The necessity for native messaging
functionality arises from a fundamental browser design philosophy: context
isolation. Modern browsers need to load and execute untrusted remote
content, while maintaining robust security guarantees. This is achieved through
sandboxed execution environments which isolate webpages, web extensions and the
local file system from each other. Most browsers also implement fine-grained
security privileges (e.g., background process management, localhost
connectivity, and local file access), all of which are strictly controlled and
enforced by the browser's security model.
Chrome goes a step further and also implements several protections to fortify itself from other desktop apps.
In 2024, Chrome introduced \textit{App Bound Encryption} for Windows,
to prevent processes running with the logged-in user privilege from 
reading sensitive user-data like cookies and
passwords~\cite{appboundencryption}.  On macOS, Chrome uses \textit{Keychain}, 
Apple's fine-grained credential storage system, which enables developers to share sensitive 
user-data with authorized applications via codesigning checks~\cite{codesigncheck}.
In 2025, Chrome
also disabled default user profile debugging via the
\texttt{--remote-debugging-port} to prevent
infostealers \cite{infostealer1,infostealer2} from stealing user data via the
Chrome Devtools Protocol (CDP).
Contemporary desktop operating systems similarly have strengthened their
security model, and implement access control mechanisms that provide isolation
between apps, and prevent cross-contamination and unauthorized data
access. For example, macOS's App Sandbox provides an isolated filesystem along
with fine-grained permission control over sensitive data locations and the
network stack. These isolation mechanisms are useful
when user workflows remain confined within individual apps or webpages, but
are a significant hindrance for cross-boundary workflows.

\textbf{Native messaging motivation.} Numerous real-world workflows necessitate
interaction between browsers and local apps. Password managers constitute a
primary example; these apps must store credentials securely on the local
machine, often within encrypted vaults, while simultaneously entering them
into web forms across disparate websites for usability reasons.
There exists a functional gap between web-facing extensions, which handle page
parsing and form population, and desktop apps, which handle secure credential storage
and retrieval. This gap also exists in Clipboard managers, which require
programmatic access to clipboard data from webpages, security tools scanning
for malicious web content, and productivity web apps requiring local file
systems, email clients, and hardware-key-based web authentication mechanisms.
The \texttt{NativeMessaging} API creates a
communication bridge between browser extensions and native apps. The
protocol was introduced by Chromium in 2013~\cite{nmorigins} and has since been
adopted by other major browsers including Firefox and Safari. At its core,
this API allows a browser extension to communicate with a native
app (also referred to as the \emph{native messaging host}) as a
separate process, and exchange structured messages over standard input
(\texttt{stdin}) and standard output (\texttt{stdout}) pipes.

\begin{figure}[!t]
    \centering
    \includegraphics[width=0.5\textwidth]{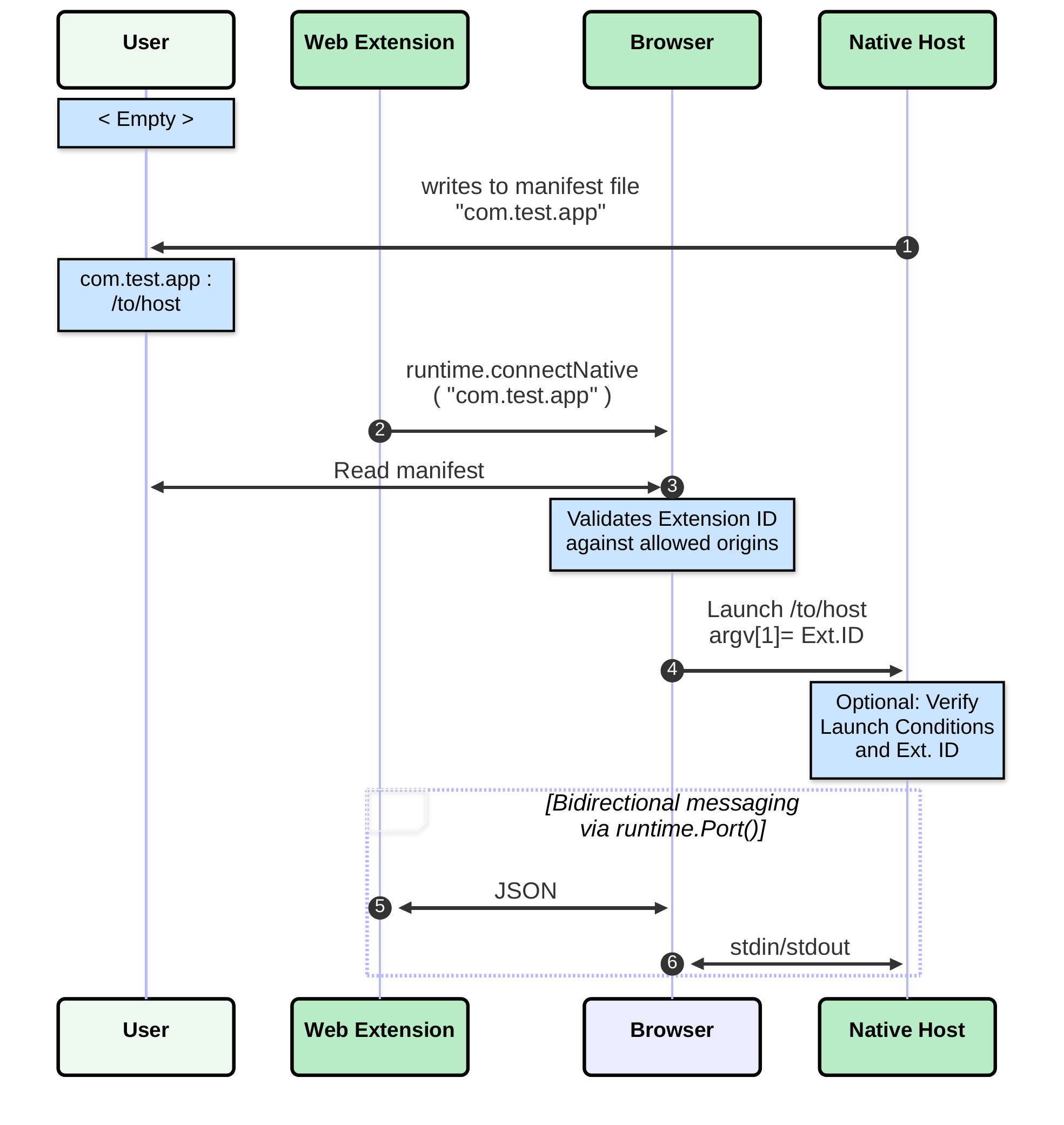}
    \caption{Overview of the native messaging protocol between the extension
    and native host, via the browser.} \label{fig:defaultnm}
\end{figure}

\textbf{Native messaging protocol design.}
Figure~\ref{fig:defaultnm} illustrates the end-to-end native messaging
workflow.

\texttt{Manifest File.} The manifest file is the central configuration that the
browser reads to determine which native host app to launch. The
location of the manifest file varies between operating systems and installation privileges. 
On macOS, user-level installations store the manifest in the \texttt{NativeMessagingHosts} 
folder within the browser's profile directory. On Windows, manifest paths are written to the registry.
The manifest is a JSON file (named in reverse DNS format) containing five fields, two of which are 
particularly significant. The \textit{path} field specifies the location of the native host executable.
Most applications we tested employed a dedicated helper binary instead of the main executable 
(see Listing~\ref{list:1p} in Appendix~\ref{sec:ManifestFileDetails}).

The \textit{allowed\_origins} field contains an allowlist of extension IDs,
which the browser checks before initiating the connection. The extension ID is
derived from the public key in the extension's manifest and is uniquely tied to
the Chrome developer's identity. Impersonating another developer's public key
is not possible on the Chrome Extension Store.  The extension also needs to
declare the \texttt{"nativeMessaging"} permission in its manifest.


\circleblack{1} The Native App starts by writing the manifest file to the
user's NativeMessagingHosts folder. To initiate the connection, the extension needs to call
\texttt{connectNative("com.1password.1password")}.

\circleblack{2} \rrow \circleblack{3} \rrow \circleblack{4}  The browser
reads the manifest file and validates the extension's origin
(chrome-extensionID) against the \path{allowed_origins} list, and launches the
native host described in the \texttt{path} parameter as a child process with
the extension's chrome origin as the first command-line argument. 
Optionally, the native host can verify this origin and conduct checks on the parent process.

\circleblack{5}\rrow \circleblack{6} 
Upon a successful connection, the browser establishes a stateless connection between the extension
and \nmh. The extension sends and receives JSON using the
\texttt{chrome.runtime.sendNativeMessage()}. The \nmh receives the
message from the browser, as each UTF-8
encoded JSON object is preceded by a 32-bit length field in native byte
order~\cite{chrome_nm_docs}. 
%
%
%
\subsection{Threat Model}
\label{sec:threatmodel}

Our research focuses on the attack surface introduced by
web browsers' \emph{native messaging} functionality.
While prior work has focused on how malicious
pages (or other malicious extensions) can target extensions' messaging
capabilities~\cite{some2019empoweb,fass2021doublex,kim2023extending,eriksson2022hardening}
we present a novel threat model which focuses on how a
\emph{malicious native app} can misuse the \nmh
capabilities to target extensions and other native 
apps installed on the user's machine.

We assume that the attacker has basic user-level permissions, i.e., a malicious
app. \emph{How} the user actually installed the malicious app (e.g., through
social engineering, visiting a malicious app store, etc.) is outside the scope
of our work. We also assume the malicious app \emph{cannot} read another
process's memory, \emph{cannot} perform system-level network monitoring or full
TLS interception, and \emph{cannot} capture keystrokes or screen content.
This is a realistic assumption for macOS, as these capabilities would require
either system-level privileges or explicit user consent prompts via TCC
(Transparency, Consent, and Control). 
In contrast, user-level apps in Windows have considerably more
capabilities, including access to other apps' files.
%

Our ultimate goal is to demonstrate attacks that \emph{elevate} a malicious
app's capabilities. For example, if the malicious app is able to read the file
system and obtain the user's private key for digitally signing documents, then
misusing native messaging to achieve the same result does not have any
additional impact. However, if the user uses a hardware key that stores
private keys, being able to man-in-the-middle native messaging communication to
stealthily replace the document being signed, then the attack significantly
enhances the malicious app's capabilities and demonstrates the risk posed by
the browser's native messaging functionality. As such, macOS is the main
focus of our analysis due to its robust built-in app isolation and integrity
mechanisms. 
Nonetheless, we also include an analysis of Windows apps for completeness;
there, the lack of strong security guarantees renders our native-messaging attacks 
mostly an \emph{additional} attack vector for local attackers.
We further discuss our threat model in \S\ref{sec:discussion} and Appendix~\ref{sec:app-local}. 

\section{Empirical Analysis of \texttt{NativeMessaging}}

We conducted an in-depth empirical analysis of browsers' Native Messaging
capabilities, by analyzing documentation, building test extensions and native
apps, and through preliminary experiments with popular extensions. All
of our experiments were conducted with Chrome v151 on macOS 26.3 and Windows 11
Education Build 10.0.26200. Firefox experiments were conducted separately with Firefox v154 on macOS 26.3.
Our analysis focused on analyzing each step of the
protocol, exploring implementation details across different operating systems,
and identifying the presence of built-in or ad-hoc endpoint identity verification
and communication encryption mechanisms, as well as identifying message
integrity and freshness checks. Next we present the flaws that we have
uncovered, which guided the design of our host-to-browser cross-boundary
attacks (\S\ref{sec:methodology}).

\textbf{Manifest Search Order.} Chromium-based browsers resolve the native host manifest by searching the 
user-level directory before falling back to the system-level 
directory~\cite{chromium_launch_context_posix, chromium_launch_context_win}.
This lookup ordering is controlled by the \texttt{allow\_user\_level\_hosts}
flag, which defaults to true~\cite{chromium_extension_prefs}.
Consequently, even when an app installs its manifest in the system-level
directory, an unprivileged attacker can shadow it by depositing 
a malicious manifest at the user-level directory under the same name. The browser
loads the user-level manifest first, never reaching the system-level
manifest, rendering the privileged installation requirement an ineffective defense.
Firefox also searches the user-level manifest location before the
system-level location across all supported platforms~\cite{firefox_nm_manifests}.


\textbf{Manifest file integrity.} Manifests in user-level directories are writable by any process running
with user privileges. On Windows, registry keys under \texttt{HKCU} require no elevation; on macOS, the user-level
directory is writable without root. An attacker can, therefore, overwrite an existing manifest, or
create a poisoned manifest (in the user-level directory) to change the \texttt{path} field to an 
attacker-controlled binary. When the extension initiates a connection, the browser launches the attacker
binary without notifying the extension or the user. 
The extension assumes it is communicating with the legitimate host, as the poisoned manifest 
retains the original \texttt{name} and \texttt{allowed\_origins}
fields. 

\textbf{Lack of Client Validation.} The Native messaging protocol lacks
built-in client authentication. The onus falls on app developers
to implement appropriate checks (e.g., the parent process' PID and source
path). macOS provides \emph{Launch Constraints}, a kernel-level system mechanism,
which allows apps to control parent processes'
identity, launch location (e.g.,~\texttt{/apps} vs \texttt{/tmp}) and launch
conditions (e.g.,~system daemon or user launched). Most apps in
our in-depth analysis (see \S\ref{sec:case-studies}) do not verify that the
browser is the parent process. 
Nonetheless, this check is not sufficient; an attacker can bypass it by launching the native host
through a headless valid Chrome binary (see \S\ref{sec:methodology}).

\textbf{Extension ID Spoofing.} While the browser can prevent a malicious
extension from initiating a native messaging connection via the \path{allowed_origins} check,
a malicious extension can spoof the legitimate extension ID by spoofing the
target extension's public key. This public key can be trivially obtained by the
attacker beforehand by examining the manifest file of the target
extension. The attacker can write a malicious extension containing the spoofed
public key in the extension manifest.  Importantly, the attacker does
\emph{not} need to get the user to install the malicious extension, nor does
the attacker need to publish it to the Chrome Store. Instead, the attacker can
create a malicious Chrome profile containing the spoofed extension and launch
it via the legitimate Chrome binary using command line flags \texttt{--headless
--user-data-dir=profile}. This bypasses both the Chrome and extension ID
checks.

\textbf{Message verification.} Since the native messaging protocol is stateless,
messages exchanged are not matched to each other.
There is also no message integrity verification built into the protocol, 
which allows a MitM attacker to tamper with the message flow.



%
%
%
%

\section{Cross-Boundary Host-to-Browser Attacks}
\label{sec:methodology}

\begin{figure}[!t]
    \centering
    \includegraphics[width=0.5\textwidth]{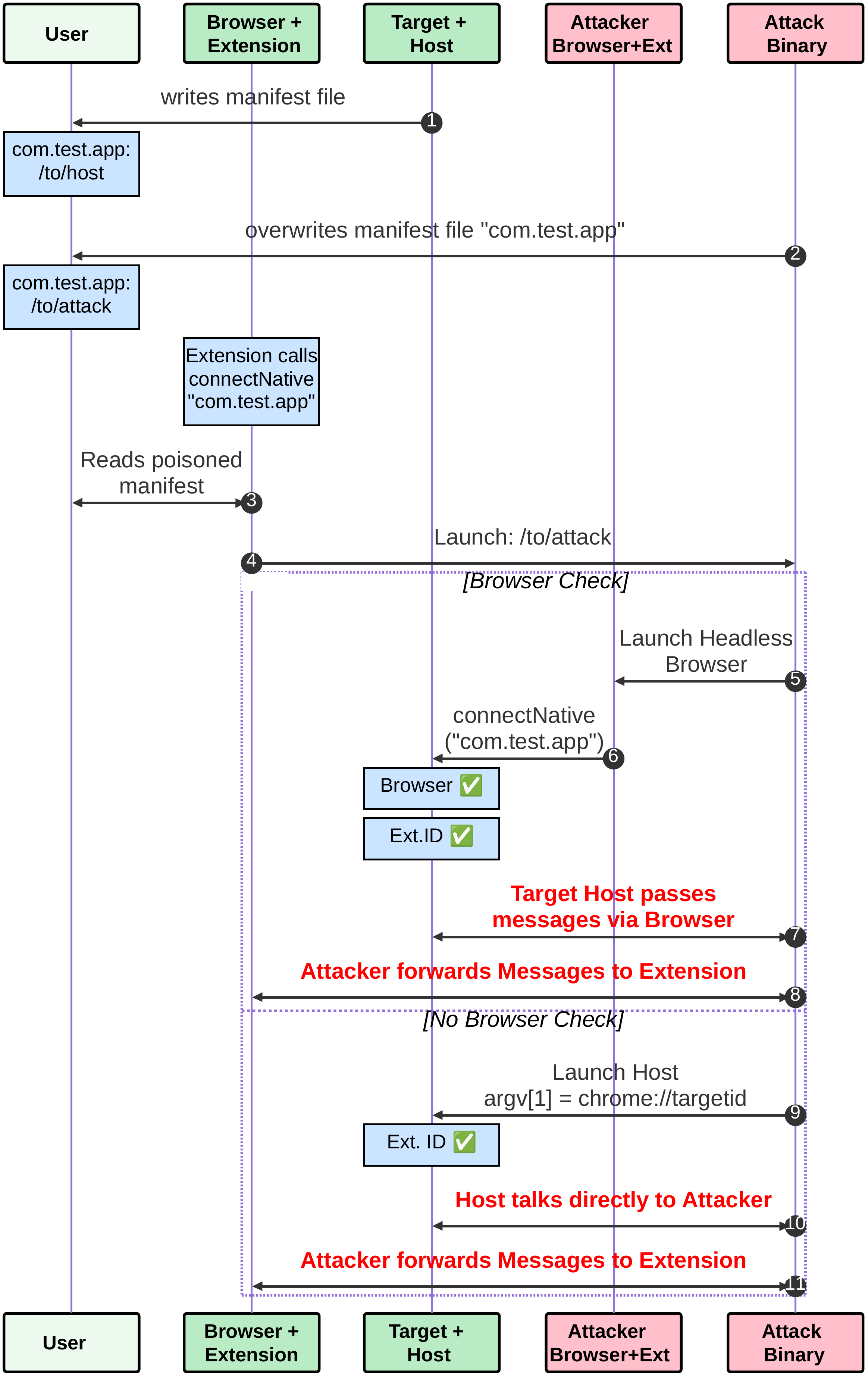}
    \caption{The attacker sits in between the Native Host and the Target Extension and can read/modify/drop any message.}
    \label{fig:bidirectional}
\end{figure}

We develop two attack variants against native messaging. 
The choice between them depends on the target's protocol data ``predictability''. \emph{Bidirectional
interception} is essentially a man-in-the-middle attack (Figure~\ref{fig:bidirectional}, detailed in this section), and is required when the extension
transmits session-specific values (e.g., account UUIDs, authentication tokens)
that the attacker must observe to maintain a coherent interception state. 
\texttt{Unidirectional interception}
(Figure~\ref{fig:unidirectional}, detailed in Appendix~\ref{sec:uniattack}) is applicable when the protocol
uses fixed endpoints with predictable payloads, wherein the attacker 
mimics one of the endpoints, conducting a
unidirectional attack.

\textbf{Manifest File.} First, we focus on the manifest.

\noindent \emph{\circleblack{1} App registration.} The target app, during its
installation process, creates an entry in the manifest file (registry key on
Windows, manifest file on macOS, see Appendix~\ref{sec:ManifestFileDetails}) that registers itself as a native messaging
host. This entry contains the app's \path{name}, the \path{path} to its
native host binary, and the \path{allowed_origins} listing the extension IDs
that are permitted to communicate with it.

\noindent \emph{\circleblack{2} Manifest poisoning.} The attacker modifies the manifest
file (or registry key in Windows) to poison the \path{path} field, redirecting
it from the legitimate native host binary to an attacker-controlled app. The
poisoned manifest retains the original fields and file name, ensuring it passes
the browser's validation checks. For apps that install their manifest in the
user-level directory, the attacker directly overwrites it, which does not require
elevated privileges. For apps that install it in the system-level directory,
the attacker instead deposits a \emph{shadow manifest} under the same host name
in the user-level directory. As described in \S\ref{sec:background}, the
browser resolves the user-level directory first, so the shadow manifest takes
effect \emph{without} the attacker needing elevated privileges.

We note that the attacker needs to ensure that the manifest
file is successfully poisoned when the extension initiates the connection. Some
apps, like 1Password and Bitwarden, overwrite the manifest file on each app
launch. So an attacker who \emph{previously} poisoned the file would not be able to
successfully intercept the connection. However, this protection can be
trivially bypassed by continuously monitoring the target manifest for changes
and re-poisoning it after each overwrite. None of the apps we tested implemented
any form of continuous monitoring of the manifest file, meaning there was no race condition
or TOCTOU issue. Moreover, we are able to intercept connections even when the target app is
already running. Furthermore, all the apps we have tested have
dedicated helper binaries functioning as the Native host; as such, the 
browser-initiated native app launch has \emph{no} discernible visual effect, such 
as a GUI app relaunch or a user prompt, that could alert cautious users.

\textbf{Extension.} The bidirectional attack assumes that the user has installed the
associated web extension in their browser profile. Unlike the Native Host, the
extension does not need to register a manifest. To detect 
the Native Host's presence, the extension simply initiates the native
messaging connection, by calling \texttt{connectNative("com.test.app")}. This
can be (i)~user-initiated, when the user explicitly triggers the extension to
connect to the native host, or (ii)~automatic, when the extension automatically
initiates contact upon startup or in response to certain events.
This Web API call will fail if there is no associated manifest registered by
the Native Host. This will also fail if the manifest does not allowlist the
extension's ID in \texttt{allowed\_origins}. Our current attack, which poisons the
manifest files, does not modify the allowed origins field of the poisoned
manifest. This allows the legitimate extension to successfully initiate the
native connection. We note that a lack of the extension registration procedure
is ``beneficial'' for the attacker, as they do not need to modify the victim's
default browser profile or any associated browser data. To prepare for the
attack, they only need to overwrite the manifest.
%
%

\circleblack{3} \rrow \circleblack{4}: Upon receiving the \texttt{connectNative()} call, Chrome reads the manifest
file (or registry key) to determine which native host to launch. Since the
manifest has been poisoned, Chrome reads the attacker-controlled \path{path}
field. Chrome spawns the attacker's native messaging host (NMH) app as a
separate child process, establishing stdin/stdout pipes for communication. The
attacker's NMH now sits between the browser and the legitimate native host.

\textbf{Browser.}
Once the attacker's native host is launched, the attacker must communicate with
the original native target. There are two distinct strategies, depending on
whether the target app validates its parent process: the attacker can easily
determine ahead of time whether each target app validates its parent browser
process.  Next we describe the scenario where the target app checks the validity
of the parent browser process; if it does not, the next phase is \circleblack{9} \rrow  \circleblack{10} \rrow  \circleblack{11}.

\circleblack{5} \rrow \circleblack{6} 
The attacker launches a headless Chrome instance configured with a custom
browser profile with a preinstalled malicious spoofed extension
(whose Chrome ID matches the target extension). This is required since browsers
pass the extension ID as the first command-line argument to the native host,
which can then verify that the correct extension initiated the connection. This
spoofed extension calls the legitimate native host via its native messaging
identifier (e.g., ``com.test.app''). To ensure that this call is not poisoned,
the attacker employs a dedicated attacker profile. Manifest files are stored
per Chrome profile; the attacker profile points to the legitimate native host
location, which can also be predetermined by the attacker during the
reconnaissance phase in a test environment. The malicious extension calls the 
legitimate target host via \texttt{connectNative("com.test.app")}. Since 
this is the attacker controlled browser profile, it has a separate \textit{NativeMessagingHosts}
folder. The manifest file used is a legitimate/non-poisoned version already prepared
by the attacker. 

\circleblack{7} \rrow \circleblack{8} The attacker's native host forwards the
intercepted message to the spoofed extension running inside the headless Chrome
instance. It also receives the message from the spoofed extension and
forwards it to the target extension, completing the interception chain.  If the
target app does not validate its parent process, the attacker can simply bypass
the browser checks we just described.

\circleblack{9} \rrow  \circleblack{10} \rrow  \circleblack{11}  
The attacker directly launches the target native host, passing a spoofed extension
ID as the first command-line argument to mimic a legitimate browser launch.
The target native host communicates directly with the attacker's binary, believing it is
interfacing with the legitimate browser extension. The attacker binary
receives the message from the target native host and forwards it to the target
extension, completing the interception flow.

\subsection{Attack Implementation Details}

\textbf{Bidirectional Attacks}
Our attack employs a long-running background process that continuously monitors
the target manifest file for modifications, re-poisoning it immediately upon
any overwrite. This process intercepts messages from the target native
host (launched by the victim's browser and extension) and routes them through
a dedicated attacker infrastructure. Specifically, the interception chain
proceeds as follows: the target extension communicates with the attacker's
native host (launched by the victim browser via the poisoned manifest); the
attacker's native host forwards messages to the background process; the
background process relays them to a malicious extension running within an
attacker-controlled browser profile; this malicious extension, configured with
a legitimate manifest pointing to the target native host, completes the
round-trip by communicating with the actual target native host. This
architecture ensures transparent MITM interception without disrupting the
app's observable behavior.

\textbf{Unidirectional Attacks}
We briefly outline two unidirectional attacks,
wherein the malicious app imitates one of the two native messaging endpoints to either
obtain sensitive information or trick the other endpoint into performing privileged actions.
We provide more details in Appendix~\ref{sec:uniattack}.

\emph{Extension imitation.} The attacker mimics
the extension and communicates directly with the target native host.
This requires only an attacker-controlled browser profile and a
malicious extension installed within that profile. The malicious extension is
configured with a valid manifest pointing to the target native host, enabling
it to launch and exchange messages with the host exactly as a legitimate
extension would. The native host processes these messages without
distinguishing them from authentic extension traffic. 

\emph{Host imitation.} The attacker mimics
the native host and receives the messages sent by the legitimate extension.
The spoofed host receives the extension's messages, extracts the
desired data, and can discard, log, or forward the messages to the
legitimate host to preserve observable behavior. This attack requires no
browser profile manipulation.

\section{\texttt{NativeMessaging} in the Wild}
\label{sec:measurement}

Here we provide a snapshot of the current
proliferation of extensions that support native messaging in the wild.

\textbf{Dataset.} We constructed our extension corpus from Palant's Chrome
Extension Manifests Dataset~\cite{palant-dataset}, a publicly maintained
archive that periodically scrapes the Chrome Web Store and stores raw
\texttt{manifest.json} files.  Using the August 14, 2026 snapshot (the most
recently available at the time of our study), we parsed manifests for 250{,}735
extensions and filtered for those declaring \texttt{nativeMessaging} in their
 \texttt{permissions} field, yielding 1{,}972 extensions with at least 1 download. 

Cumulatively, these extensions have approximately 1.07 billion
Chrome installations.  We find that half of the extensions have a total of
approximately 1K downloads, indicating low adoption among users.  However, the
top 5\% of the extensions account for hundreds of millions of downloads
cumulatively, demonstrating the widespread adoption of extensions that request
the necessary permission for access to the browser API that supports native
messaging functionality. While one unidirectional variant of our attack does not even require
the presence of the extension, our
measurement shows that native messaging extensions are widely present across
user machines, further highlighting the importance of an in-depth security
analysis of this browser API. Nonetheless, in practice, even more users may be
vulnerable to our attacks, as they may have installed desktop apps with native
messaging support without installing the corresponding browser extension.

\begin{figure}[!t]
    \centering
    \includegraphics[width=0.35\textwidth]{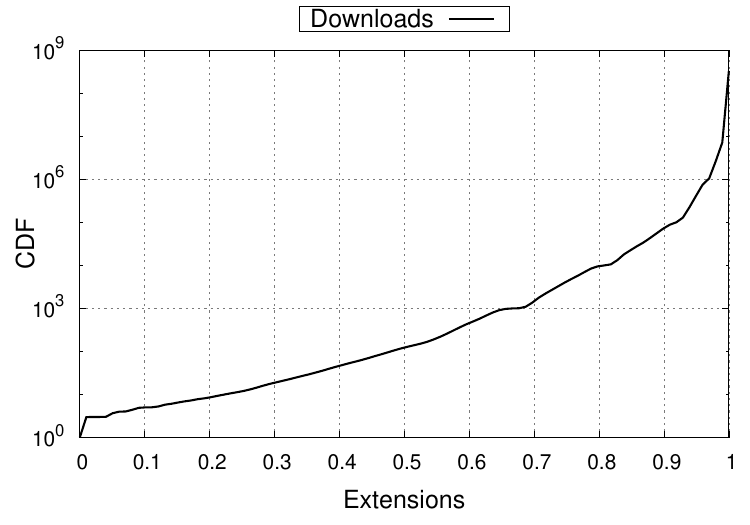}
    \caption{Extensions that support native messaging.}
    \label{fig:cdf}
\end{figure}

\section{Attack Use Cases}
\label{sec:case-studies}

\begin{table*}[!t]
\centering
\footnotesize
\rowcolors{2}{white}{gray!12}
\begin{tabularx}{\textwidth}{@{} l r @{\hspace{0.8em}} c @{\hspace{0.8em}} c @{\hspace{1em}} X @{}}
\toprule
\textbf{Extension}
  & \rotatebox{45}{\textbf{Installs}}
  & \rotatebox{45}{\textbf{Platform}}
  & \rotatebox{45}{\shortstack{\textbf{Attack}\\\textbf{Variant}}}
  & \textbf{Attacker Capabilities} \\
\midrule

Adobe Acrobat              & 329M  & \faApple\,\faWindows            & $\leftrightarrow$\                              & Cloud documents exfiltration; PDF output silently substituted with attacker-controlled content \\
Launcher for Google Drive  & 91M   & \faApple\,\faWindows            & $\rightarrow$\,/\,$\leftrightarrow$\            & Hijack file launch by victim; Steal file by routing to attacker-controlled app. \\
Chrome Remote Desktop      & 38M   & \faApple\,\faWindows            & $\rightarrow$\,/\,$\leftrightarrow$\            & Access victim screen and interact via keyboard and mouse. Can also bypass TCC on macOS.\\
Cisco Webex                & 22M   & \faApple\,\faWindows            & $\rightarrow$\,/\,$\leftrightarrow$\            & Meeting hijacked to attacker session; victim's mic and camera exposed \\
Claude in Chrome           & 15M   & \faApple\,\faWindows            & $\rightarrow$\,/\,$\leftrightarrow$\            & Arbitrary JS execution and browser actions in victim's session \\
Foxit PDF Creator          & 8M    & \faApple$^\dagger$\,\faWindows  & $\rightarrow$\,/\,$\leftrightarrow$\            & Attacker-chosen URL converted to PDF for victim \\
WPS PDF                    & 6M    & \faApple$^\dagger$\,\faWindows  & $\rightarrow$\,/\,$\leftrightarrow$\            & Persistent WPS cloud session hijacked; victim's documents replaced \\
Bitwarden                  & 7M    & \faApple                        & $\leftrightarrow$\                              & Master key extracted in plaintext via MitM key exchange \\
iCloud Passwords           & 6M    & \faApple\,\faWindows            & \xmark                                          & Not exploitable \\
1Password                  & 7M    & \faApple\,\faWindows            & $\rightarrow$\,/\,$\leftrightarrow$\            & Vault secret key, decryption keys and victim's account details silently exfiltrated \\
Web PKI                    & 4M    & \faApple\,\faWindows            & $\rightarrow$\,/\,$\leftrightarrow$\            & Victim impersonation, legally-binding signing of attacker document\\ 
Web Signer                 & 2M    & \faApple\,\faWindows            & $\rightarrow$\,/\,$\leftrightarrow$\            & Victim impersonation, legally-binding signing of attacker document\\ 
Web eID                    & 1M    & \faApple\,\faWindows            & $\rightarrow$\,/\,$\leftrightarrow$\            & Impersonation on any Web eID-protected service without key extraction \\
KeePassXC-Browser          & 0.8M  & \faApple\,\faWindows            & $\rightarrow$\,/\,$\leftrightarrow$\            & Credentials extracted in plaintext via MitM key exchange \\
RoboForm                   & 0.6M  & \faApple\,\faWindows            & $\rightarrow$\,/\,$\leftrightarrow$\            & Full vault exfiltrated: credentials, credit cards, passport, and PII \\
UiPath                     & 0.05M & \faApple\,\faWindows            & $\leftrightarrow$\                              & Arbitrary JS injection; unintended and attacker controlled browser automation \\
Automation 360             & 0.1M  & \faApple\,\faWindows            & $\leftrightarrow$\                              & Unintended and attacker controlled browser automation \\
\bottomrule
\end{tabularx}
\caption{Exploitation results across all evaluated extensions.
    \textbf{Platform:} operating system(s) on which the extension was tested.
    \textbf{Attack Variant:} $\rightarrow$~unidirectional (attacker controls
    one side of the channel); $\leftrightarrow$~bidirectional (full MitM);
    {\xmark}~application was not exploitable. $^\dagger$On macOS only possible
    attack is host imitation (no native host installed).}
\label{tab:exploitation}
\end{table*}

Here, we present a series of concrete attack demonstrations that highlight the
severe security ramifications of our host-to-browser cross-boundary attacks.
For a subset of our collected extensions, we describe the attack surface,
detail the exploitation chain, and characterize what an attacker can achieve.
Table~\ref{tab:exploitation} summarizes our findings, and
Table~\ref{tab:versions} (Appendix~\ref{sec:targetdetails}) provides app
versions.

\textbf{Target selection.} From our main collection of extensions
(\S\ref{sec:measurement}), we selected 17 for in-depth testing, chosen to
balance install-count coverage, category diversity, and sensitivity of the
data they process. We prioritized popular extensions with large user
bases because they are more likely to have security protections in
place (as is typically the case with popular websites~\cite{chen2016longitudinal,vansteenhuyse2025one}).
We further focus on extensions that handle sensitive user data and mediate
security-critical workflows, since their compromise poses a severe threat.  Our
selected targets also span diverse categories to demonstrate that the security
risks introduced by native messaging are not isolated to a single domain, but
reflect a broader weakness in browser-host communication mechanisms. The
vulnerable extensions account for $\sim$532 million installations across
the Chrome ecosystem (Table~\ref{tab:exploitation}). All the 
target apps were tested on both macOS and Windows, except for Bitwarden as we 
did not have a Windows device with a biometric sensor. Finally,
while other browsers also support extensions and \texttt{NativeMessaging},
we focus our analysis on Chrome as it is the most prevalent browser.
We have, however, verified that Firefox is also vulnerable by replicating our
attacks on a subset of the targeted applications: Web eID, Web PKI, 1Password,
KeePassXC, Bitwarden, and RoboForm.

 \textbf{Web eID.} The European Union's eID ecosystem underpins a wide range of
security-critical digital services, including citizen authentication, legally
binding electronic signatures, and access to government resources, making it a
foundational component of Europe's emerging digital infrastructure. Web eID is the open-source 
framework developed to enable ID card-based
authentication and digital signing on the web, developed by Estonia's Information System Authority 
(RIA)~\cite{ria_web_eid}, and is currently supported by numerous EU member countries~\cite{id_ee_web_eid} 
(with support already underway in many others).
The Web eID extension uses native messaging to access hardware interfaces such as
\texttt{PC/SC} via its native host. The app exchanges APDU commands with the
ID card's chip which stores the authentication and signing
keys~\cite{cybernetica_2020}.
During authentication, the extension
forwards a server-issued challenge and origin information to the native host,
which instructs the smart card to produce a signature over the challenge via
the \texttt{authenticate} endpoint. 
Similarly, when signing a document, the extension forwards a server-provided
document hash through the \texttt{sign} command, and the native host requests
the card to sign the supplied digest~\cite{web_eid_app}.
The attacker is able to modify these calls in transit,
resulting in the user unwittingly signing an attacker's challenge nonce or document.
We tested our bidirectional attack variant using the official Web eID app~\cite{web_eid_eu} and a 
Windows TPM-backed Virtual Smart Card (VSC) with a self-signed root certificate.
For the relying party, we deployed the official
\texttt{web-eid-authtoken-validation-php} reference
implementation~\cite{web_eid_php_example}, which explicitly supports replacing
the default Estonian CA bundle with any certificate set.
To initiate the attack, the attacker first visits the Web eID-protected service
on a separate machine; without the need for a smart card, a phantom native host installed
via manifest poisoning intercepts the \texttt{authenticate} command and
captures the server-issued nonce $N_2$, which is staged on the victim's
machine. When the victim subsequently authenticates on any Web eID service, the
MITM binary intercepts the \texttt{authenticate} command carrying the victim's
nonce $N_1$. In the simplest variant, the MITM replaces $N_1$ with $N_2$: the
victim enters their PIN once, the attacker obtains a token signed over $N_2$,
but the victim's own authentication fails. In the stealthier variant, the MITM
sequentially spawns a second \texttt{web-eid-app} process and drives it to sign
$N_2$ under the attacker's origin (PIN entry~1), then forwards $N_1$ to the
legitimate host (PIN entry~2), completing the victim's authentication normally.
The resulting token --- bearing the victim's certificate and a valid signature
over $N_2$ --- is returned to the attacker, who submits it to their own session
and is authenticated as the victim, without the private key ever leaving the 
smart card's TPM. We note that the two prompts are unlikely to alert users;
a user study on FIDO2 security found that participants were not alarmed by attacks
that required two consecutive authentication prompts each requiring a physical tap,
attributing them to improper key taps or a glitch in the protocol~\cite{yadav2023security}.

We note that a unidirectional Extension Imitation attack variant is also
possible.  However, it is less stealthy as the native host presents a PIN
prompt to the user.  As such, in practice, this can alert more cautious users 
if no preceding user action has been performed to account for the prompt.
%
%

\textbf{Web Signer and Web PKI.} Web Signer, developed by Softplan, 
is the digital signing component
used by Brazil's e-SAJ electronic judicial filing portal for signing documents.
Its users include over 320,000 registered lawyers for whom Web Signer installation is a mandatory prerequisite for
electronic filing~\cite{esaj_tjsp}. Brazilian legal practice distinguishes two ICP-Brasil certificate
classes: A1 certificates are software-stored, while A3 certificates reside on hardware tokens such as
smart cards or USB cryptographic devices; for legally-binding judicial filings, A3 tokens are
mandated. Web PKI is identical to  Web Signer platform: the
underlying native host binary is identical, the native messaging protocol is the same, and the
extension manifest accepts both the Web Signer and Web PKI extension identifiers. And, similarly, 
the A3 token's hardware boundary prevents key
extraction, but native messaging controls what is signed.

The \nmh interface exposes three security-critical commands.
\texttt{listCertificates} returns all certificates installed on the device,
each identified by a thumbprint\footnote{A thumbprint is the base64 encoding of
the SHA-256 hash of the certificate's DER-encoded bytes.} alongside full
metadata. \texttt{authorizeSignatures} carries the thumbprint of the
certificate the user selects and is the only point of user interaction in the
signing flow: the native host presents a consent dialog, and upon approval
returns an authorization permitting subsequent signing operations with that
certificate. \texttt{signHash} carries a server-computed document hash and the
chosen certificate thumbprint; the native host silently signs the hash with the
certificate's private key and returns the raw signature bytes -- no further
user interaction occurs. To assemble a complete signed document, the server
requires two values submitted by the browser: the raw signature bytes from
\texttt{signHash} and \texttt{certThumb}, the signer's base64-encoded DER
certificate,\footnote{DER (Distinguished Encoding Rules) is the standard binary
serialization format for X.509 certificates; \texttt{certThumb} carries its
base64 encoding.} both visible in the browser-to-server HTTP traffic. The
native host additionally supports a \texttt{readCertificate} command, which
accepts a thumbprint and returns the full base64-encoded DER certificate; this
command is not part of Chrome's normal signing flow but is fully supported by
the host. The raw signature bytes and 
\texttt{certThumb} allow the attacker to reconstruct the signed document.


For our evaluation, we used the official Web PKI demonstration
portal~\cite{webpki_demo} as the relying party. To replicate the signing
process without a valid ICP-Brasil certificate, we generated a self-signed
certificate. When the victim initiates document signing, the MITM app
intercepts the \texttt{signHash} command and performs a dual-sign transparently
within the single authorization the user already granted. It first
forwards the legitimate document hash $h_1$ to the native host, captures the
authentic signature, and returns it to Chrome, so the victim's document is
signed normally. It then immediately issues a second \texttt{signHash} carrying
the attacker's target hash $h_2$, and captures the resulting spoofed
signature. Separately, when the \texttt{listCertificates} response passes
through the pipe, the MITM silently injects a \texttt{readCertificate} command,
invisible to Chrome and extracts the certificate's \texttt{certThumb}. The
attacker then assembles a complete, cryptographically-valid signed document
using the signed $h_2$ and \texttt{certThumb} from the server.

\textbf{1Password.} 1Password is a highly popular password manager which employs native messaging for
user convenience. It stores user secrets and passwords in a vault which is shared across
all app instances via 1Password's servers. Once the desktop app is unlocked by the user,
it uses native messaging to automatically unlock the browser extension.
This model introduces a critical flaw.

The API design allows an extension to continuously query the Native Host for the vault's state, via
the \texttt{NmRequestAccounts} API call, which does not require any unique key
or secret to initiate. If the vault is unlocked, the API call
returns the user's full account details, including email address, the fixed
secret key (recovery key), and the current decryption keys~\cite{1pass_mal_proc}.
These secrets allow an attacker to replicate the user's entire vault on their own machine. 
It is also important to note that vault-unlocking is part of the normal user flow,
and the user needs to unlock it to obtain any passwords. Once the vault is unlocked, this API
returns all the aforementioned information to the attacker without any user prompt, or visual indication.

Since 1Password enables native message functionality by default, this attack
succeeds even if the user has not installed the browser extension.  When native
messaging is enabled, any app running with standard user privileges can
\emph{invisibly} query the user's secret key and current decryption keys via
this endpoint.
1Password emphasizes its security design, storing the secret key in macOS's Secure Enclave
and Windows' protected app space to mitigate memory scanning~\cite{1pass_device_unlock_sec}.
Its security white paper explicitly excludes memory scanning and trojanized apps from its
threat model, claiming resistance only against ``less powerful malware'' operating under
user-level privileges.
However all these protections are rendered void due to the exposed
unauthenticated native messaging endpoint. 

\textbf{Roboform.}
The desktop app enables native messaging by default and encourages users to install the
browser extension during installation.
The API exposed over native messaging allows the extension to query files within the app's data
directory via the \texttt{rf-api-request} endpoint. This wrapper API exposes individual functions
such as \texttt{getfile}, \texttt{list}, and dedicated functions for extracting
the user's login key and password. These allow the extension to
enumerate the vault contents and read sensitive information including credit card numbers and passport information.
The extension can also directly retrieve the user's login credentials in plaintext using these endpoints. 
However, similar to 1Password, the app must be unlocked by the user before any data can be
accessed. Once the user unlocks RoboForm to use it, the native messaging
endpoint immediately becomes available, allowing any unprivileged app
to query and extract the vault contents without requiring any user prompt,
notification, or a separate authentication step. 

\textbf{Bitwarden, KeePassXC-Browser.} Unlike the previous examples, both of these password managers 
disable native messaging communication by default.  Bitwarden restricts native
messaging to biometric authentication and requires users to enable it in both
application and extension. KeePassXC-browser similarly requires explicit opt-in.
Both apps also use asymmetric encryption to secure their native messaging channel.
Bitwarden uses RSA-2048 OAEP key exchange to establish an AES-256-CBC channel with HMAC-SHA256 integrity; KeePassXC-Browser uses
TweetNaCl's \texttt{crypto\_box} construction and nonce-based replay 
prevention~\cite{keepassxc_protocol}. Both mechanisms are cryptographically sound when implemented correctly.
However, since the attacker sits between the extension and the app, they can perform independent key exchanges with both parties,
establishing two separate encrypted sessions and decrypt and re-encrypt all traffic transparently.

The Bitwarden API exposes a simple, \texttt{getBiometricStatusForUser}, and
\texttt{unlockWithBiometricsForUser} endpoint which allows the extension to
fetch the user key from the native host. However, these API calls
require the \texttt{appID} and \texttt{userID} parameters, which can only be
obtained via a bidirectional attack.  Additionally, the attacker still requires
access to the encrypted vault contents to make use of the intercepted key.  For
KeePassXC-Browser, the \texttt{test-associate} check does not obstruct the MITM: the attacker relays
the stored identification key inside its own re-encrypted session,
providing no proof of private key possession --- a weakness KeePassXC's own
protocol specification concedes~\cite{keepassxc_protocol}, and one
independently identified in a public security analysis of the
protocol~\cite{keepassxc_github_issue}. The \texttt{get-logins} response,
carrying credentials in plaintext within the decrypted session can be
intercepted transparently. 
KeePassXC \emph{is} susceptible to extension imitation attacks, however, a
fresh \texttt{associate} request produces a visible dialog in KeePassXC with
no corresponding browser action to explain it, and each subsequent
\texttt{get-logins} call independently triggers the per-URL access
prompt. The Bidirectional attack is stealthy within the
normal app flow and leaves no such indicator.
%

\textbf{iCloud Passwords.}
This is the only password manager we evaluated which
was not vulnerable to the attack despite using the native messaging protocol.
The app's native messaging manifest is symlinked to
a location protected by System Integrity Protection (SIP), meaning that
even an attacker with root privileges cannot modify or replace the manifest
file. Even placing a malicious
manifest in the user's directory does not override the system manifest.
As a result, iCloud Passwords cannot be attacked.
\textbf{PDF-signing tools.} We also test several PDF tools under the bidirectional attack model.
The Acrobat Reader Chrome extension uses native messaging to open cloud-stored
documents in the desktop app, and to convert web pages to PDF.
Document-opening requests transmit raw PDF contents to the native host,
allowing an attacker controlling the communication channel to obtain copies of
documents retrieved from the user's Adobe cloud account without interacting
with Adobe's infrastructure. During webpage conversion, the extension forwards
the page DOM and embedded resources to the host, which generates the resulting
PDF. By modifying these inputs, an attacker can silently inject arbitrary
content into the generated document without any user-visible indication.
WPS PDF and Foxit PDF Creator also rely on native messaging to
transfer document and conversion requests from the browser to a desktop 
app. In WPS PDF, document-opening requests contain the raw document
contents, allowing an attacker controlling the communication channel to
exfiltrate or modify files before they are opened by the user. The extension
additionally exposes a single-use WPS SSO token during its automatic login
workflow, enabling the attacker to obtain a fully authenticated session to the
victim's cloud account without user interaction. Foxit PDF Creator similarly
forwards conversion requests to its native host when generating PDFs from
browser tabs. By modifying the transmitted URL, an attacker can cause Foxit to
create a PDF of attacker-controlled content while reporting successful
completion to the user. Neither app registers a manifest in macOS,
so data exfiltration is achievable via Host Imitation: a shadow manifest redirects
the extension's connection to an attacker binary, bypassing the need
to intercept an existing channel.

\textbf{Launcher for Google Drive. } This enables users to open cloud-stored
files from Google Drive with their preferred local desktop app.  The native
host exposes two primary endpoints: \texttt{apps\_for\_file}, which queries the
host for all apps capable of handling a given file type (e.g., Adobe Acrobat
for PDFs), and \texttt{open\_file}, which launches the selected app with the
target document. File contents are never transmitted over the
native messaging channel; instead, Google-assigned \texttt{file\_id}
identifiers are exchanged which are resolved independently against Google's
servers to retrieve the underlying content. An attacker can use the
Bidirectional attack to intercept and modify the \texttt{open\_file} request
and substitute the legitimate app identifier with a malicious one. A
stealthier attack variant involves the attacker injecting an additional poisoned
\texttt{open\_file} request alongside the legitimate one to deter suspicion. 

\textbf{Chrome Remote Desktop.} This tool 
allows users to connect to remote clients via their Google Account. 
The native host operates with elevated privileges, including screen capture, input forwarding
using XMPP with a custom (\texttt{$<$UUID$>$@chromoting.gserviceaccount.com}) account.
While the native host exposes multiple API endpoints, 
\texttt{startDaemon} allows the extension to associate the native host daemon
with a new account. This endpoint is typically used when a user registers the
machine and requires an OAuth token, but can be initiated by the extension at
any time. An attacker using a throwaway Google account can exploit this
endpoint by initiating the device registration process with an
attacker-controlled extension which talks directly to the target native host.
While this does require user consent via a system dialog box, the dialog
is generated by the legitimate Chrome remote desktop app and simply states that
``Chrome Remote Desktop Host wants to make changes to your PC'',
which is the same message shown when the app is installing an update.
This attack achieves full device takeover allowing the attacker to
also compromise other connected clients and bypass TCC prompts on macOS.

\textbf{Claude in Chrome.} Claude Desktop provides a companion extension to
interact with the browser on the user's behalf~\cite{claudeinchrome}.  The
extension uses the Devtools Protocol to instrument the browser and
provides a range of automation capabilities, including navigation,
click actions, and form filling.  The native app can send requests to the
extension over native messaging, and the extension executes the actions within
the browser and returns the results. The API is structured around the
\textit{tool\_request} and \textit{tool\_response} message types.  The
extension exposes a suite of tools, including \textit{navigate},
\textit{tabs\_context}, \textit{file\_upload}, \textit{tabs\_create},
\textit{left\_click}, \textit{key}, and \textit{scroll},
among others.  Furthermore, the extension emits system prompts via
\texttt{<system-reminder>} tags, which include operational suggestions such as
the use of batched tool invocations.  Interestingly, the extension and the
native app do not typically use native messaging. Instead, each component
independently establishes a WebSocket connection (WSS) to the Anthropic server
which bridges their communication. Nevertheless, the extension retains native
messaging functionality, which presents an attack surface that could be
exploited by malicious native apps.

Claude in Chrome is vulnerable to both Host imitation and Bidirectional
Attacks.  While Chrome does display a visible indicator- "Claude" is
debugging- when Claude is active, this indicator is present regardless of
attacker presence. An attacker can perform any of the aforementioned browser
actions, including direct JavaScript execution on any tab being debugged by
Claude in Chrome.  Since the extension is designed to be used with logged-in
user accounts, the capability allows the attacker to stealthily obtain tokens,
cookies and other sensitive user data directly from the logged-in context. We
note that other UI-based actions (e.g., click, navigate) might be less
stealthy; however, when Claude is configured with the ``act without asking'' mode,
the extension does not prompt the user for permission before executing
sensitive operations, making such actions entirely seamless.  Moreover, an attacker may inject content into the
\texttt{<system-reminder>} prompt intended for the native host, to manipulate 
its behavior.

\textbf{Cisco Webex.} 
extension uses native messaging to open meeting links in the Webex desktop app.
It is functionally similar to URL-deeplinking, without Chrome's user consent
dialog box~\cite{deeplink}.  When the user navigates to a Webex link, the
extension sends an unencrypted \texttt{launch\_meeting} request to the native
host, containing the meeting link and password.  The API call also includes
meeting and desktop app parameters.
An attacker can use the bidirectional modification attack to manipulate meeting
parameters forcing the victim to launch an attacker-controlled meeting.  The
attacker has control over participant metadata, display names, and other
profile attributes, which can facilitate social engineering attacks often used
in AI-powered deepfake scams~\cite{crypto_fake_meeting}. The attacker can also
modify critical app configuration settings, including thirdPartyLogin options,
telemetry collection URLs, and app update URL paths.  It is also possible to
conduct a unidirectional Extension Imitation attack on the Webex native host,
but it is less stealthy since the app prompts the user for consent when
launching a meeting.

\section{Discussion}
\label{sec:discussion}

Here we discuss certain aspects of our attacks and experimental findings,
and the underlying issues, which inform our mitigation strategies discussed
in the following section.

\textbf{Trust boundary crossing.} By design, the native messaging protocol
crosses trust boundaries. Browser extensions operate with elevated privileges
granted by users, including the ability to read and modify web page content,
intercept requests and responses, and interact with web APIs. Native
apps, meanwhile, operate with access to system resources. When an
extension communicates with a native host, it bridges the browser's isolated
execution environment with that of native apps. Trust boundaries operate
on the assumption that the data and content within each boundary are trustworthy;
however, trust is not inherently transferable across boundaries. Web extensions
receive explicit user-approved access to web content at installation time, while
native host apps receive explicit or implicit access to system resources.
Incorrect assumptions about trust transferability can lead to the omission of
necessary security controls: native apps may read web content they were never
intended to access, while extensions may read system resources outside the
browser sandbox, each operating under the false assumption that the other's
trust domain confers equivalent authorization.

\textbf{Client validation in message passing.} Cross-boundary
communication mechanisms typically rely on a \textit{root of trust} to
authenticate clients and authorize access across security boundaries. For
example, macOS XPC binds client identity to developer certificates and audit
tokens~\cite{koboldxpc}, Android Binder uses kernel-verified UIDs, PIDs, and
permissions, and Windows named pipes enforce kernel-controlled ACLs. RPC and
distributed systems similarly use client certificates, authentication tokens,
or shared secrets. Without client validation, privileged services can become
confused deputies, allowing less-privileged processes to misuse their
authority~\cite{koboldxpc,wang2022iservice,androidvulns}. Prior work has
demonstrated such attacks against XPC and Binder services, as well as broader
cases where unprivileged apps gain access to resources intended for
privileged contexts~\cite{koboldxpc,xing2015unauthorized,lin2025chekhov}.

Native Messaging breaks from this established security model. It
provides no comparable client validation: native hosts have no reliable means
to authenticate the extension initiating a connection.
Consequently, any process capable
of establishing a Native Messaging connection can impersonate a legitimate
extension and forge messages. This weakness is compounded by Chromium's default
user-level manifest resolution, which checks user-level registrations before
system-level ones~\cite{chromium_launch_context_posix,chromium_launch_context_win}. An
unprivileged process can therefore register a shadow manifest that preempts a
legitimate host. Together, 
these weaknesses allow an unprivileged app to impersonate an
extension, intercept and modify its communication, and forge messages to
security-sensitive native hosts.

\textbf{Local Attacker Threat Model.} 
We assume an unprivileged app with no elevated privileges or permissions.
We emphasize that user-level code
execution is not equivalent to unrestricted host compromise. 
This distinction is particularly important on modern macOS, where OS-enforced
protections restrict such apps from accessing other apps' data, capturing
keyboard or screen input, or freely interacting with other processes. These
isolation guarantees substantially constrain the capabilities of an
unprivileged app, but Native Messaging provides a mechanism for circumventing
these boundaries: by communicating with trusted native apps through the
browser, an attacker can indirectly access security-sensitive functionality
that would otherwise remain inaccessible. This does not imply that macOS
apps are impervious to local compromise, as app-specific
vulnerabilities or other avenues may still expose protected data.
The impact of native messaging is more limited on Windows, where
user-level apps have substantially broader capabilities. The security
implications, therefore, depend on the isolation guarantees provided by the
underlying OS: on platforms with strong isolation, Native Messaging can bridge
security boundaries that would otherwise constrain an unprivileged app, whereas
on platforms with weaker isolation, it instead represents an alternative attack
vector.

\textbf{Native-Messaging Attack Advantages on macOS.} Native Messaging can
substantially amplify an unprivileged local attacker's capabilities by
providing access to functionality otherwise constrained by the OS's app
isolation and data-protection mechanisms. To concretely quantify these
capabilities, we analyzed all target apps under a non-native-messaging
local attacker model, examining what an attacker could achieve using only data
and functionality directly exposed by the apps and their extensions.
Our analysis shows that most security-critical attacks in our
evaluation (i.e., against 1Password, Bitwarden, Roboform, Web PKI, Web Signer, Web eID)
\emph{cannot be reproduced} by a non-native-messaging local
attacker. In these cases, Native Messaging is not merely an alternative attack
path: it provides the capabilities necessary to cross OS-enforced app
boundaries and access otherwise protected functionality and data.

In other cases, Native Messaging substantially amplifies
attacker capabilities. In Cisco Webex, a local attacker can recover meeting
details, but Native Messaging enables modifying and controlling meeting
launches, facilitating the social-engineering attack from \S\ref{sec:case-studies}. Similarly, Native
Messaging enables modification of Chrome Remote Desktop's configuration, turning
otherwise passive access into our cross-registration attack from \S\ref{sec:case-studies}.
For Launcher for Google Drive, a local attacker can recover Drive file IDs but
cannot access the corresponding files; Native Messaging enables their stealthy
exfiltration. 

Access to user folders like \textit{Desktop} and \textit{Documents} on macOS is
gated behind TCC, which prevents a local attacker from accessing them unless
the user grants the malicious app that privilege.  As such a local attacker
would not be able to read and modify such documents without leveraging the
native messaging techniques against such apps as WPS, and Adobe Acrobat.
Interestingly, in the case of Adobe Acrobat we found that if the user logs
into their Adobe account, an access token is exposed by the extension which
allows a non-native-messaging local attacker to access the user's account.
However, if the user is not logged in, the local attacker cannot attack Adobe
Acrobat, whereas the native messaging attack can still achieve the attacks
detailed in \S\ref{sec:case-studies}. Unlike Acrobat, the WPS PDF extension caches
previously opened PDFs in local storage, which are accessible to the local attacker; 
however, native messaging exposes
an SSO token enabling persistent cloud session hijacking.

Finally, while an insecurely stored OAuth token in Claude Desktop allows local
attackers to access the user's account and conversation-history, Native
Messaging enables arbitrary JavaScript injection into the user's Chrome Default
Profile. Essentially, this enables attacker-controlled code execution in the
context of websites the user is authenticated to, enabling actions that are
otherwise unavailable to the local attacker.  Moreover, UiPath exposes a
refresh token which can be used by a local attacker to hijack the associated user
account, leading to workflow compromise. However, native messaging allows the
attacker to also inject JavaScript in the browser's context, significantly augmenting 
their capabilities. On the other hand, KeepassXC's extension and Foxit's native host leak 
secrets to local attackers that enable account compromise.

Overall, these results demonstrate that Native Messaging is a distinct
capability-amplifying attack surface, not simply another path to
vulnerabilities already exposed by apps. Its impact is particularly
significant on secure operating systems such as macOS: in most
consequential cases, OS protections prevent the attack entirely without Native
Messaging, while in others Native Messaging transforms limited local access
into active exploitation.


\textbf{App Sandboxing.} Both macOS and Windows support sandboxing for
isolating apps from other apps and sensitive data, and minimizing the impact of
installing a malicious app. We evaluated whether app sandboxing prevents
our attacks, and found that an attacker operating within the macOS App Sandbox or Windows
AppContainer is generally unable to exploit Native Messaging due to
insufficient privileges; however, sandbox exceptions
can allow the attacker to overcome this restriction.\footnote{While our threat model
assumes a non-privileged local attacker, we further discuss sandbox exceptions
in Appendix~\ref{sec:sandbox:exceptions}, for a more complete presentation of
the sandboxing process and its use in real-world apps.}

More importantly, unlike mobile platforms, sandboxing is not mandatory for desktop
apps: both macOS and Windows permit users to install non-sandboxed
apps outside their official stores. This is particularly relevant for
macOS, where many app developers actually avoid the
official app store (due to revenue loss, review delays, and incompatibility
with sandboxing restrictions~\cite{mac_dev_survey2}). Moreover, only 20\% of
surveyed developers distribute exclusively through the Mac App
Store~\cite{mac_dev_survey}, while 66\% of users install apps from outside
it~\cite{mac_user_survey}; 76\% of these users rely on developer sites.
Indeed, only four apps in our study are available through the Mac App Store,
and the available 1Password version is severely outdated (v7.9.11, released
07/2023). On Windows, ten apps are available in the store, but seven
use \texttt{runFullTrust} and three install Win32 executables, leaving none
effectively sandboxed. Overall, attackers are not constrained to distributing
malicious apps through official stores; as long as users routinely
install desktop software directly from developer sites, attackers can
distribute unsandboxed apps through the same established channel,
making store-enforced sandboxing insufficient as a general defense.

\section{Hardening Native Messaging}
\label{sec:countermeasure}

Preventing our host-to-browser cross-boundary attacks requires changes
along three dimensions, which are complementary, each targeting a distinct attacker
capability and each closes a gap others leave open. The first restricts which processes can register a native host
manifest. The second restricts which processes can invoke the native host
binary. The third removes trust from the channel entirely, and requires 
setting up encryption at the app layer.

\textbf{Manifest poisoning.} 
Since shadow manifest attacks require only user-level privileges, they can be
neutralized by changing a single Chromium default: setting
\texttt{kNativeMessagingUserLevelHosts} to \texttt{false}
would restrict host resolution to system-level registrations, requiring
administrator/root privileges for manifest
manipulation~\cite{chromium_extension_prefs}. As an upstream Chromium change,
this would propagate to Chromium-based browsers, 
though vendors can independently adopt it. The main
compatibility cost is that user-level native host installations would require
elevated installers.  Until an upstream change is made, administrators on
managed on-premises devices can enforce the equivalent restriction via the
\texttt{NativeMessagingUserLevelHosts} policy through Group Policy on Windows
or MDM on macOS~\cite{chromium_nm_userlevel_policy,
chrome_enterprise_nm_policy}; on user machines, the user can switch to the
administrator profile and apply the same restriction directly via the system
registry or managed preferences.  For existing user-level manifests, browsers
can additionally monitor registration paths and block connections when their
on-disk values change; extensions can similarly validate the resolved binary
path against a value recorded at installation.


\textbf{Process invocation.} Restricting manifest writes does not prevent an attacker from invoking the
legitimate native host from a non-browser parent. Patching this requires
kernel-enforced validation of the caller's identity before the host begins
execution. On macOS, Launch Constraints~\cite{apple_launch_constraints_docs}
provide this
mechanism, enforced by AMFI at execution time. A constraint blob embedded in the
binary's code signature describes the conditions the parent process must satisfy;
if the expression evaluates to false, the \texttt{exec} call returns
an error and no process is created. The blob is part of the signed
data, so modifying it also invalidates the signature. On Windows,
no kernel-level equivalent exists. A native host can attempt an application-level
check but it's bypassable.

Only iCloud Passwords embedded a parent constraint out of all
the macOS native hosts we tested. It enumerates approximately thirty browsers
by signing an identifier and team identifier; spawning the host from any
unlisted parent is rejected by the kernel before a single instruction executes.
No third-party native host in our dataset embeds equivalent constraints.  These
two measures together --- restricted manifest resolution and kernel-enforced
parent validation, prevent an unprivileged attacker from substituting a
malicious binary and from invoking the legitimate binary from an unauthorized
parent. Nonetheless, the headless-browser attack variant remains possible; a
legitimate browser launched with an attacker-controlled profile satisfies any
parent-identity constraint, because the browser binary is unmodified. A
protocol-level mitigation narrows this gap: the browser could supply the
native host with a cryptographic hash of the calling extension alongside the
extension ID, allowing the host to reject connections from spoofed extensions
that present the correct ID but not the expected code. This strategy is standard in IPC systems: the
IPC layer attaches unmodifiable caller identity to each message, which the
receiver then authenticates. macOS XPC services
derive the caller's code-signing identity from the kernel-attached audit
token~\cite{apple_seccode_audittoken}, and Android Binder services combine the
kernel-supplied caller UID with cryptographic package-signature
verification~\cite{android_hassigningcertificate}; Chromium's own security
documentation identifies the latter as ``the only reliable mechanism by which
you can authenticate your peer''~\cite{chromium_android_ipc}. Native messaging
lacks this property; we argue for the need to extend the same pattern to the
browser-native boundary, with the browser playing the role the kernel plays
for process-to-process IPC.


\textbf{Encrypted communication.} Since protocol-level changes require browser
vendor adoption, apps can instead independently harden the channel.
Both Bitwarden and KeePassXC-Browser attempt in-band encryption, and
both fail for the same structural reason: key establishment runs over the
channel the attacker already controls. The attacker performs independent
exchanges with each endpoint, establishes two encrypted sessions, and relays
decrypted traffic between them. The fix is to move key establishment
out-of-band. The extension and native host should each authenticate to a
shared backend over TLS, and receive a per-session symmetric key;
traffic encrypted under this key cannot be read or forged by an attacker on the
native channel, because the key never traverses
it~\cite{diffie1992authentication,dierks2008transport}. Where a shared backend is
unavailable, a Password-Authenticated Key Exchange (PAKE)~\cite{bellovin1992encrypted} as adopted by iCloud Passwords,
achieves this property locally.

\section{Related Work}
\label{sec:related}

\textbf{Browser extensions.} Several prior studies have examined the security
risk and attack surface of browser extensions. Pantelaios et al.~\cite{pantelaios2020you} showed that benign extensions may transition
to malicious behavior through obfuscated code updates, 
while Wang et al.~\cite{mali_ext_crypto} highlighted the
proliferation of cryptocurrency-themed extensions that facilitated malvertising
and credential theft. Eriksson et al.~\cite{eriksson2022hardening} identified
inter-extension attack vectors, including password stealing via DOM
manipulation, traffic redirection for search query interception, and
cross-extension XSS through history-poisoning.
The closest work to ours is by Some~\cite{some2019empoweb}, which showed that while
browser extensions and web apps are executed in separate contexts, they
can interact through message-passing channels that malicious pages can exploit
for various attacks, including bypassing SOP and reading user cookies. 
While these studies consistently demonstrate that browser extensions represent
a high-value target for attackers, the scope of threats was restricted to
interactions \emph{within} the browser, i.e., extension-to-webpage or
extension-to-extension. Attacks that \emph{misuse} extensions from \emph{outside}
the browser, which we show in our work, have not been previously explored.
Previous external attacks against extensions focused on 
\emph{replacing} installed extensions with malicious versions~\cite{ter2008enhancing}.

\textbf{Desktop apps.} Security weaknesses in modern desktop
apps have been studied across a variety of platforms and deployment
models. Xiao et al.~\cite{xiao2022understanding} demonstrated that unsafe
cross-context interactions in desktop apps can be exploited to achieve
Remote Code Execution (RCE), potentially leading to full system compromise;
their work instrumented the V8 engine to identify and restrict cross-context
control flows, with particular emphasis on Electron's IPC mechanisms. 
Carettoni~\cite{electronegativity} provided one of the earliest systematic examinations
of Electron security and introduced Electronegativity, a static-analysis tool
for detecting common security misconfigurations in Electron apps.
Paloscia
et al.~\cite{paloscia2025lost} highlighted the challenges of porting web
apps to desktop environments, showing that security assumptions
inherited from the web often fail in native execution contexts and introduce
new attack surfaces. 
Ali et al.~\cite{inspectron} further
investigated popular Electron-based apps
via dynamic analysis and identified multiple insecure
design patterns that could facilitate RCE attacks.
Bui et al. \cite{bui2018man} looked at MiTM attacks in IPC communications between non-privileged
apps and their privileged helpers,
and found socket hijacking vulnerabilities in password
managers; 
however, no vulnerabilities in Native Messaging were identified.
Ahmadpanah et al.~\cite{ahmadpanah2021sandtrap} analyzed
trigger-action platforms such as IFTTT, Zapier, and the open-source Node-RED,
discovering vulnerabilities that enable cross-app code execution and
man-in-the-middle (MitM) attacks ranging from data exfiltration to full
platform takeover. Wang et al.~\cite{wang2024rootfree} demonstrated how weak
privilege enforcement on Windows can be abused to launch cross-platform attacks
from desktop environments against mobile super-apps, including WeChat.

Collectively, these studies reveal that desktop apps frequently expose
complex attack surfaces arising from privileged execution environments,
cross-app interactions, and the integration of web technologies.
However, prior work has not investigated the attack surface
introduced by native messaging, a standardized inter-process communication
mechanism that enables direct, bidirectional communication between browser
extensions and native desktop apps, with existing (or planned) integration
in extremely sensitive government-related workflows. 

\textbf{Password managers.} Prior work has studied credential theft against
password managers across web and mobile environments. Silver et
al.~\cite{silver2014password} showed that malicious websites and network
adversaries can exploit browser autofill, while Lin et al.~\cite{lin2020fill}
and Fu et al.~\cite{fu2024leaky} demonstrated that hidden forms and concealed
inputs can trigger unintended autofill and credential exfiltration. Fratantonio
et al.~\cite{fratantonio2017cloak} identified abuses of Android Accessibility
services to access password-manager data, and Gangwal et
al.~\cite{gangwal2023autospill} showed that credentials autofilled into
WebViews may be exposed to the hosting app. Anliker et al.~\cite{309832}
examined phishing attacks against password-manager extensions, while Infantino
et al.~\cite{vaultraider} identified UI-based attacks against major desktop
password managers.
Unlike these attacks, ours do not rely on user-facing mechanisms such as
autofill or UI manipulation, or on elevated permissions. Instead, we exploit
Native Messaging to extract credentials invisibly, without user prompts or
password-manager interaction.

Recently, Reguła~\cite{woj_keepassxc} explored security and isolation mechanisms on macOS with a
focus on novel local attacks on password managers. While most of the attacks rely on
memory injection, the attack on KeePassXC focuses on insecure communication between
the browser extension and the main app. However, unlike our Native Messaging approach, this attack
extracts plaintext credentials from the extension's local storage.
Hoffman and Morgan~\cite{defcon} previously presented attacks
against the 1Password macOS app, including the use of native messaging. 
Critically, their attack~\cite{robinhoodpoc} involves dynamic instrumentation 
via FRIDA to force Chrome to \texttt{fork} and \texttt{exec} the native host
and hook the \texttt{stdin/stdout} pipes. \emph{By default}, macOS prevents
userland processes from hooking and instrumenting other apps. Enabling this 
requires the user to reboot the entire system and disable System Integrity Protection (SIP)~\cite{sipdisable},
which is a \emph{significantly} higher-privileged attack (with limited practicality)
compared to the \emph{unprivileged} host app in our threat model.
Moreover, our work presents more complete
threat modelling (including the bidirectional and host imitation attacks),
a more systematic analysis of the \texttt{NativeMessaging} pipeline
with new findings (e.g., manifest poisoning), and diverse
case studies that characterize the prevalence, and broader security
implications. 

\section{Conclusion}
\label{sec:conclusion}

Our research presents the first systematic security analysis 
of browsers' native messaging functionality.
Our in-depth exploration of the native messaging protocol and functionality 
uncovered fundamental flaws that can be abused by \emph{any} unprivileged
app running on the user's machine. Guided by our findings,
we present two variants of our host-to-browser cross-boundary attack through
a series of concrete attack demonstrations with severe real-world implications.
Our extensive evaluation shows that an unprivileged app can abuse
native messaging capabilities exposed by critical services,
enabling adversaries to subvert security-critical workflows
without requiring elevated privileges or user-granted permissions. Importantly,
the security impact is greatest on platforms that otherwise provide strong
app isolation; native messaging can undermine OS-enforced boundaries by
providing a low-privilege app with indirect access to capabilities and
data that the OS would otherwise restrict.
The widespread presence of these
vulnerabilities across diverse app domains highlights the systemic
nature of the problem. We, therefore, propose mitigations for the entities
involved in the native messaging workflow to harden the communication pipeline
and restore the security boundaries that native messaging can otherwise
circumvent.


\section*{Acknowledgments}\label{sec:acks}

We would like to thank the anonymous reviewers and shepherd for their valuable feedback.
This  work  was supported  by the National Science Foundation under grants
CNS-2211574 and CNS-2143363.  Any opinions, findings, conclusions, or
recommendations expressed herein are those of the authors, and do not
necessarily reflect those of the NSF.

\section*{Competing Interests} 
The authors have no competing financial or non-financial interests.

\bibliography{paper}

\appendices
\section{Manifest File  Details}
\label{sec:ManifestFileDetails}

Here we provide additional details about the native messaging manifest file.

\subsection{Manifest File Location}

\emph{Windows:} During installation, apps create a registry key at one of these paths:
\begin{itemize}
    \item \footnotesize \path{HKEY_CURRENT_USER\Software\Google\Chrome\NativeMessagingHosts}
    \item \footnotesize \path{HKEY_CURRENT_USER\Software\Wow6432Node\Google\Chrome\NativeMessagingHosts}
    \item \footnotesize \path{HKEY_LOCAL_MACHINE\SOFTWARE\Google\Chrome\NativeMessagingHosts}
\end{itemize}
\normalsize The default value of the key contains an absolute path to the Native Messaging manifest.

\emph{macOS:} The manifest file is stored at two locations depending on user privileges:
\begin{itemize}
    \item \footnotesize \path{~/Library/Application Support/Google/Chrome/NativeMessagingHosts} (current user's location)
    \item \footnotesize \path{/Library/Google/Chrome/NativeMessagingHosts} (system-wide location)
\end{itemize}
\normalsize
%
\normalsize
Most of the apps we tested install their manifest file in the user-level
directory. Some security-sensitive apps install in the system-wide directory, since it requires
root privileges for macOS and administrator privileges for Windows.

\subsection{Example manifest file}

\begin{listing}
\begin{minted}[fontsize=\tiny,
    breaklines,
    bgcolor=gray!5,
    frame=lines,
    rulecolor=\color{black},
    framesep=6pt
    ]{json}
    {"name": "com.1password.1password",
  "description": "1Password BrowserSupport",
  "path": "/apps/1Password.app/Contents/Library/LoginItems/1Password Browser Helper.app/Contents/MacOS/1Password-BrowserSupport",
  "type": "stdio",
  "allowed_origins": [
    "chrome-extension://aeblfdkhhhdcdjpifhhbdiojplfjncoa/"
  ]}
\end{minted}
    \caption{1Password manifest file.}
\label{list:1p}
\end{listing}

Listing~\ref{list:1p} shows 1Password's manifest file, which points to the dedicated helper binary for handling native messaging.

\section{Unidirectional Attacks}
\label{sec:uniattack}

\begin{figure}[!t]
    \centering
    \includegraphics[width=0.5\textwidth]{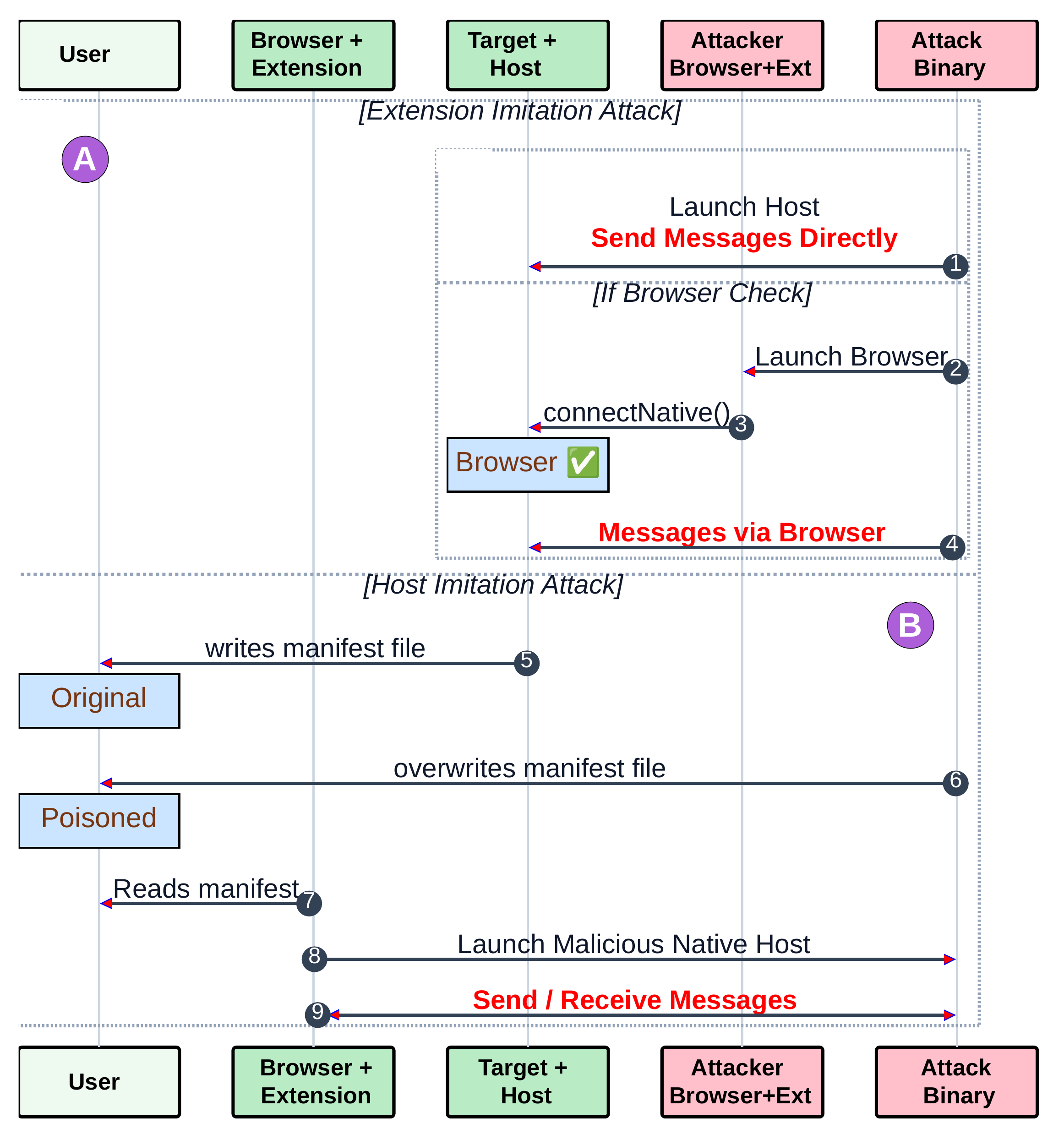}
    \caption{(A) Extension Imitation Attack: The attacker only attacks the
    Native Host by pretending to be the Extension. (B) Host Imitation Attack:
    The attacker only attacks the Target Extension by pretending to be the
    Native Host.} \label{fig:unidirectional}
\end{figure}

Figure~\ref{fig:unidirectional} details the unidirectional version of our
cross-boundary host-to-browser attack.  The benefit of this attack is that the
adversary can also target users that only have one of the two target components
installed on their machine (i.e.,~only the extension or only the native app).
The diagram presents two distinct variants of the unidirectional attack:

\textbf{\circlepurple{A} Extension Imitation Attack:}  The attacker communicates directly with the \nm host by impersonating
a legitimate extension. The user does not need to have the legitimate extension for this attack to work.

\circleblack{1} : 
If the \nm host doesn't check their parent process, the attacker can
directly spawn the native host as a child process. Extension ID checks implemented on the host
can be bypassed by passing the spoofed ID as the first command-line argument when launching it.

\circleblack{2} \rrow \circleblack{3} \rrow \circleblack{4} : 
If the \nm host performs parent process validation,
the attacker launches a headless Chrome instance containing a malicious spoofed extension.
Extension ID matches that of the legitimate extension. The extension then launches the \nm host via the
legitimate Chrome binary thereby bypassing the parent process validation check.
The spoofed extension's ID is passed via Chrome to the \nm host bypassing any host-side checks.

In both cases, the attacker is able to communicate directly with the host.

\textbf{\circlepurple{B} Host Imitation Attack:}  The Attacker communicates directly with the browser extension
by impersonating
a legitimate \nm host.  This attack works even if no \nm host is ever installed by the user.

\circleblack{5} \rrow \circleblack{6}: The attacker either creates or overwrites the target app's  manifest with
a poisoned \texttt{path} pointing to attacker-controlled binary.

\circleblack{7} \rrow \circleblack{8} \rrow \circleblack{9}: Once the extension call \texttt{connectNative}, the browser 
launches the attacker binary which can then communicate with the extension via the browser.

%

\section{Tested Extensions and Native Hosts}
\label{sec:targetdetails}

Table~\ref{tab:versions} provides information and the specific version tested for each component.

\begin{table}[!t]
\centering
\caption{Tested versions of extensions and native hosts. {\small\faChrome}~extension version; {\small\faTerminal}~native host version.}
\label{tab:versions}
\scriptsize
\rowcolors{3}{white}{gray!12}
\resizebox{\linewidth}{!}{
\begin{tabular}{@{}l @{\hspace{0.5em}} c@{\hspace{0.3em}}c @{\hspace{0.5em}} c@{\hspace{0.3em}}c@{}}
\toprule
\multirow{2}{*}{}
& \multicolumn{2}{c}{\textbf{Windows}}
& \multicolumn{2}{c}{\textbf{macOS}} \\
\cmidrule(lr){2-3} \cmidrule(lr){4-5}
\rowcolor{white}  & \faChrome & \faTerminal
& \faChrome & \faTerminal \\
\midrule
Adobe Acrobat          & 26.8.2.1 & 26.1.21771.0 & 26.5.2.3 & 7.0 \\
App Launcher for Drive & 3.10 & 126.0.5.0 & 3.10 & 126.0 \\
Chrome Remote Desktop  & 2.1 & 149.0.7827.18 & 2.1 & 147.0.7727.3 \\
Cisco Webex            & 2.0.4 & 10056.4506.2025.1214 & 2.0.4 & 46.5.0.35006 \\
Claude in Chrome       & 1.0.75 & 1.0.0 & 1.0.75 & 1.11187.4 \\
Foxit PDF Creator      & 12.1.0.0 & 12.1.0.25862 & 12.1.0.0 & --- \\
WPS PDF                & 1.4.0 & 2025.1.1.10517 & 1.0.0.55 & --- \\
iCloud Passwords       & 3.3.0 & 15.9.0 & 3.2.0 & 2.4 \\
Bitwarden              & --- & --- & 2026.2.0 & 2026.2.1 \\
1Password              & 8.12.22.17 & 8.12.22.17 & 8.12.22.17 & 8.12.22.17 \\
Web PKI                & 2.17.5 & 2.14.2.0 & 2.17.5 & 2.14.2 \\
Web Signer             & 2.17.1 & 2.11.1.0 & 2.17.1 & 2.12.1 \\
Web eID                & 2.3.0 & 2.9.0.927 & 2.3.0 & 2.8.0 \\
RoboForm               & 9.9.6.0 & 9.9.4.6 & 9.9.6.0 & 9.9.4 \\
KeePassXC-Browser      & 1.10.2 & 2.7.12 & 1.10.2 & 2.7.12 \\
UiPath                 & 26.10.1 & 25.10.178.58789 & 26.10.1 & 8.0.12.6609 \\
Automation 360         & 4.3.0.0 & 1.5.0.253321140 & 4.3.2.0 & --- \\
\bottomrule
\end{tabular}
  }
\end{table}

\section{Local Attacker Capabilities}
\label{sec:app-local}
Next we provide additional details about the capabilities of a local attacker.

\textbf{Filesystem Access.} Windows does not prevent user-level
processes from accessing other apps' data. macOS is far more
restrictive; TCC protects sensitive resources, and App Containers (tied to
Apple Developer Team IDs and enforced by SIP) secure data even for
non-sandboxed apps. Moreover, native hosts can also store sensitive data in
the macOS keychain which is also not accessible by other apps.

\textbf{Native Host Tampering.} Local Attackers may attempt to modify other user apps (e.g., password managers)
in an attempt to steal data or keyboard input.
They can achieve this by tampering with the target's app's binary file on disk or dynamically injecting code at runtime.
Windows does not provide the same level of memory protection as macOS, so on
Windows local attackers can tamper with other native apps through process hollowing or \texttt{dll} injection. 
In contrast, macOS by default verifies the app bundle integrity at runtime and allows app developers to enforce
runtime restrictions preventing any processes, even root-owned ones, from modifying or reading app memory.

\textbf{Extension Tampering.} Chrome implements safeguards to mitigate
the damage which can be caused by a local attacker.  For example, a local
attacker cannot directly modify the extension files to achieve JavaScript execution in
the context of the Default User profile. Chrome implements an integrity check
on Extensions downloaded via the Chrome Web Store which prevents a local
attacker from trivially modifying the extension code stored on disk. Any
modification to the extension code triggers a corruption warning from Chrome
and disables the extension altogether. The user is presented with a choice to
repair the extension which reverts any modifications made by the attacker.
Similarly, an attacker cannot add a new extension by directly modifying
Chrome's Default profile. Prior Chrome versions allowed loading
arbitrary extensions from disk via \texttt{--load-extension} flag but this was
deprecated due to misuse. Adding an extension now requires enabling
developer mode via GUI and manually loading in the extension from disk, which
is beyond the capabilities of a user-privileged attacker on macOS.

However, while native messaging is intended to offload sensitive
functionality and data to the native apps (e.g., the app storing the password
vault and the extension simply autofilling it in the webpage) certain extensions may
still store sensitive data. If the extension stores it in an
unencrypted form, that would be readable by other local apps. In practice,
security-critical apps typically do not store sensitive data unencrypted.
E.g., password manager extensions that include the password vault, prompt the
user for their Master Password, and pass that through a key derivation function
to calculate the key used to decrypt sensitive data in memory, thereby
preventing local attackers from accessing it. 

\section{Sandboxing Exceptions}
\label{sec:sandbox:exceptions}

Since strict sandboxing often prevents legitimate apps from operating
normally, both Windows and macOS implement a mechanism for exceptions for
necessary capabilities which can be granted to apps.

For example, in macOS \texttt{com.apple.security.temporary-exception.files.home-relative-path.read-write},
is required to access the \texttt{NativeMessagingHosts} directory. 
Apps distributed via the App Store which use native messaging contain exceptions,
giving the sandboxed apps access to native messaging manifest directories across all major browsers.
Malicious apps published through the App store that have been provided with this exception are able to overwrite legitimate manifests and create fake ones,
allowing them to mount Host Imitation Attacks against target extensions.
However, mounting Extension Imitation or the full MITM attack would require more sandbox exceptions or broader permissions.
The Windows App store is far more lax, allowing apps to completely forgo sandboxing, by requesting the
\texttt{runFullTrust} capability. 7 of the apps in our experiments were granted this capability.
In summary, while attackers can request exceptions to overcome the limitations of sandboxing, in practice attackers
can simply choose to avoid the official app stores given desktop users' tendency to install apps directly from websites.


\section{Ethics Considerations}
\label{sec:ethics}

\textbf{Purpose and Impact of Research.} The native messaging protocol, which
enables bidirectional communication between web browsers and native host applications,
remains a significantly understudied component of modern browser architectures.
Despite a lack of academic scrutiny, the protocol is widely deployed across
popular browsers and serves as a critical infrastructure layer for numerous
productivity, security, and utility applications. Our research is motivated
by the imperative to secure user data in this under-protected communication
channel. We find that the native messaging protocol suffers from significant
flaws that, if left unaddressed, enable severe attacks. Given the massive user
base of applications leveraging this protocol, our findings carry significant
real-world impact for end-user privacy and security.

\textbf{Experimental Conduct.} All experiments described in this paper
were conducted exclusively on the researchers’ own machines, using isolated
test browser profiles and synthetic or personally owned data. No actual end
users were targeted, contacted, or otherwise implicated in our experimental
procedures. All data collection and analysis was performed in a controlled,
environment to ensure that no third-party data was accessed without explicit
authorization.

\textbf{Beneficence.} Our research involves several stakeholder groups, 
each of whom stands to benefit from the findings and recommendations presented 
in this work, as we discuss next. \textit{Browser Vendors.} Our empirical evaluation 
focuses primarily on Google Chrome. Nonetheless, the design flaws we identify 
may not be isolated to a single browser, as they likely extend to other Chromium-based 
browsers and potentially other browsers as well. Our findings reveal significant 
architectural shortcomings in the native messaging security model that warrant a 
comprehensive redesign. As such, our research is beneficial to browser vendors as 
it provides the blueprint for them to incorporate additional mechanisms to mitigate 
our attacks. \textit{Application Developers.} Our research highlights a previously unknown 
attack vector that enables unauthorized access to sensitive user data through the 
native messaging protocol. As application developers bear responsibility for securing 
their applications against such threats, we have initiated disclosure with the developers 
of the applications evaluated in this study. Similarly, application developers 
that want to incorporate native messaging, will benefit from our research as it 
will allow them to secure their apps against the attacks we present. \textit{End Users.} 
By exposing the security deficiencies inherent in the native messaging protocol, 
our research aims to raise awareness of the risks embedded in the products that 
end users rely upon daily. We believe that informed users are better positioned 
to make decisions about which applications to install. Our findings may empower 
users to exercise greater caution when interacting with applications that leverage 
native messaging, and to advocate for stronger privacy protections from vendors and 
developers.

\textbf{Responsible Disclosure.} We adhere to responsible disclosure practices
throughout this research. 
For each app, we parsed their official websites or public repositories to identify the appropriate disclosure mechanisms.
From browser vendors, Google, Mozilla, Microsoft Edge, Opera, and Vivaldi were contacted. Google and
Mozilla explicitly consider user-level local attackers outside their browser
threat models;\footnote{\url{https://issues.chromium.org/issues/40080979}} Google closed all reports on this basis, including those for
Chrome Remote Desktop and Launcher for Google Drive.
Edge attributed the issue to upstream Chromium, and Opera
and Vivaldi did not commit to changes. These responses reflect the
long-standing practice of treating user-level local attackers as out of scope.
However, this assumption is increasingly outdated as OS security has evolved,
particularly on macOS, with stronger app integrity guarantees and
restrictions on user-level apps’ access to sensitive resources such as
other apps’ files, keyboard input, and screen data. Treating local
execution as equivalent to unrestricted host compromise overlooks these
boundaries and risks leaving cross-app attack vectors, such as those
enabled by native messaging, systematically understudied.

Feedback from the targeted apps varied considerably. The Estonian
Information System Authority (RIA) (administrator of Web eID) acknowledged the
attack surface and after considerable communication proposed a vendor-neutral
security framework as a longer-term desired direction.  Several vendors
confirmed the attacks and implemented or pursued mitigations: Foxit deployed
HMAC-SHA256 message integrity and anti-replay protections; Anthropic confirmed
the issue was previously reported and is tracked for remediation; and RoboForm
is exploring potential changes.  1Password, Bitwarden, and KeePassXC also
engaged with the reports and assessed the implications for their threat models
but did not commit to any changes. Cisco (Webex) was determined
that their implementations did not cross a security boundary, while Adobe has
not provided a final response. Lacuna Software and Automation
Anywhere acknowledged receipt without substantive responses; UiPath and WPS PDF
did not respond. These outcomes demonstrate that disclosure can prompt
meaningful improvements, but also highlight the difficulty of securing
native-messaging channels at the app level. Doing so requires apps to correctly
implement and maintain message integrity, authentication, replay protection,
and appropriate origin/browser validation across diverse deployment
environments. We argue that native messaging should instead provide these
guarantees by design, avoiding the need for each app to independently secure a
complex browser-to-host communication channel.


\textbf{Decision to publish.} We believe that the benefits of publishing our 
work significantly outweigh any potential risk. As critical workflows leveraging 
native messaging are emerging (including citizen-government processes in the EU), 
it is imperative that vendors are aware of the risks that they face. Moreover, we 
propose mitigations steps that can assist developers and vendors in deploying 
appropriate mechanisms for securing their systems.

\newpage
\section{Meta-Review}
The following meta-review was prepared by the program committee for the 2027
IEEE Symposium on Security and Privacy (S\&P) as part of the review process as
detailed in the call for papers.

\subsection{Summary}
This paper presents a new privilege escalation attack that abuses the Native 
Messaging API to target communication between browser extensions and their 
native host counterparts under a local-attacker threat model. The paper 
demonstrates the feasibility of man-in-the-middle (MITM) and confused deputy 
attacks against Chrome extensions and unsandboxed macOS applications. Through 
an empirical study of real-world extensions, the paper shows that these attacks 
affect widely deployed applications across several security-sensitive domains.

\subsection{Scientific Contributions}
\begin{itemize}
    \item Identified an Impactful Vulnerability
    \item Provides a Valuable Step Forward in an Established Field
\end{itemize}

\subsection{Reasons for Acceptance}
\begin{enumerate}
    \item This paper identifies an impactful vulnerability that is exposed
    by the Native Messaging API ecosystem. Through an analysis of 17 Chrome
    extensions and their unsandboxed macOS native hosts, the authors find that
    16 are vulnerable, spanning more than 500 million installations, although
    this count includes non-macOS installations. The affected applications
    include government verification applications, password managers, and a
    popular AI platform.
    \item This paper provides a valuable step forward in an established field
    by examining the Native Messaging API, which has received comparatively
    little attention despite extensive prior research on browser extensions.
    The paper presents the first systematic evaluation of this mechanism and
    thus advances an established line of browser security research.
\end{enumerate}

\subsection{Noteworthy Concerns}
\begin{enumerate}
    \item The PC noted that the paper demonstrates meaningful privilege
    escalation primarily on macOS. While the attack was also replicated on
    Windows, the assumed local attacker already has substantial privileges there,
    limiting the additional impact demonstrated by the attack. Further, given
    the attack is feasible even without the browser being present, the evaluation
    may not capture the full attack surface.
    \item The reviewers also agreed that the threat model is restricted by
    requiring unsandboxed native applications. This may limit the applicability
    of the attack, since applications downloaded from OS vendor-monitored stores
    do not fall within this threat model.
    \item The threat model already grants the local attacker considerable
    capabilities. As such, a more systematic presentation of the exact additional
    capabilities enabled by this attack, compared with what an attacker with the
    assumed local access can already achieve, would strengthen the paper.
\end{enumerate}

\section{Response to the Meta-Review}
\begin{enumerate}
    \item Indeed, our attack is significantly more impactful in macOS,
    due to the various OS-level defenses present in macOS that considerably
    limit the capabilities of user-level non-privileged processes (see discussion
    in Section 7). Moreover, as our experiments demonstrate, attackers can exploit
    the NativeMessaging API against certain targets even if the user has not
    installed the browser extension or the native app (see Table 1). As such,
    we agree that other attack techniques may be possible using native messaging;
    nonetheless, our paper presents an extensive, in-depth analysis of the NativeMessaging
    API, and demonstrates a powerful novel attack vector.
    \item We broadly agree that the threat model is somewhat restricted by requiring
    the attacker to be unsandboxed. However, we contend that this has limited real
    world implications given the popularity of unofficial distribution channels in
    desktop environments. As we discuss in Section 7, surveys have found that macOS
    app developers tend to avoid the app store due to revenue loss, review delays,
    and incompatibility with sandboxing restrictions. In fact, only 4 out of the 17
    apps in our evaluation were available via the Mac App Store despite their popularity,
    while all of them were available outside the app store through the vendors’ official
    websites.
    \item We also agree that a local attacker possesses certain capabilities
    by default. Thus for some target applications, the native messaging channel
    provides an alternative vector rather than the only one. However, our empirical
    testing of non-native-messaging local attacker capabilities showed that numerous
    attacks  demonstrated in the paper cannot be mounted without exploiting the Native
    Messaging channel. The difference between local attacker capabilities and the
    importance of our native messaging attacks are significantly more pronounced in
    macOS, as the numerous built-in OS-level defenses constitute significant barriers
    for local attacks and largely constrain their capabilities.
\end{enumerate}

\end{document}